# *Holding and amplifying electromagnetic waves with temporal non-Foster metastructures*

*Victor Pacheco-Peña[1†], Yasaman Kiasat[2†‡], Diego M. Solís[2,3], Brian Edwards[2], and Nader Engheta[2*]*

[1]*School of Mathematics, Statistics and Physics, Newcastle University, Newcastle Upon Tyne, NE1 7RU, United Kingdom*
[2]*Department of Electrical and Systems Engineering, University of Pennsylvania, Philadelphia, PA 19104, USA*
[3] *Departamento de Tecnología de los Computadores y de las Comunicaciones, University of Extremadura, 10003 Cáceres, Spain*

*\*email: engheta@seas.upenn.edu*

## Abstract

We introduce a mechanism that can both hold and amplify electromagnetic waves by rapidly changing the permittivity of the medium during the wave travel from a positive to a dispersionless (i.e. non-Foster) negative value and then back again. The underlying physics behind this phenomenon is theoretically explored by considering a plane wave in an unbounded medium. Interestingly, we show that a rapid positive-to-negative temporal change of $\varepsilon(t)$ causes the propagation of the wave to stop (observed by a *frozen* phase in time) while the amplitude of the frozen field exponentially grows. Stepping the permittivity back to the original (or a new) positive value will cause the wave to *thaw* and resume propagation with the original (or the new) frequency, respectively. We numerically study the case of dipole radiation in such time-varying non-Foster structures. As a possible implementation, we propose a parallel plate waveguide platform loaded with time-dependent media emulating parallel lumped non-Foster negative capacitors. Such non-Foster time-varying structures may open new venues in controlling and manipulating wave-matter interaction.

† These authors contributed equally to this work.

‡ Present address for YK: 210 Locust St, 16E, Philadelphia, Pennsylvania 19106, USA, yasaman.kiassat@gmail.com

## Introduction

Slowing down wave propagation has been of great interest to the scientific community. Multiple approaches have been demonstrated to reduce the speed of light and spatially trap it, such as ultracold atomic gases[1], optical resonators[2], photonic crystals[3], and wave interference in disordered media[4] inspired by the Anderson localization in electron transport[5]. Also of interest is light amplification which has been achieved using several effects such as gain and nonlinear media. On a different front, the idea of four-dimensional (4D) metamaterials in which the material not only varies in three-dimensional space (3D) but also in time (1D) has seen growing interest in recent years[6–15]. The time-dependent electromagnetic platforms have a long history, dating back to the middle part of the 20[th] century[16,17]. In past studies, the relative permittivity (ε) of the medium was rapidly changed in time from one positive to another positive value (both greater than unity). This effectively acted as the temporal analog of the spatial interface between two semi-infinite materials with different ε, resulting in a forward (FW) (i.e., time-refraction) and a backward (BW) (i.e., time-reflection) waves[18–20]. This analogy between spatial and temporal interfaces has been recently exploited to explore various interesting phenomena[13,21]. For instance, the notion of photonic time crystals (PTC) has been developed wherein the dielectric permittivity is modulated in time between two positive values, functioning as the temporal counterpart of spatial photonic crystals[22]. In this context, Lyubarov *et al*. theoretically show that this can produce light amplification and propose the exciting idea of nonresonant PTC lasers.[23]

Inspired by the opportunities offered by spatiotemporal metamaterials for wave-matter interaction, here we introduce a new mechanism that can hold *and* amplify a propagating electromagnetic wave. Fundamentally different from PTCs, the effect is based on a single temporal transition between a dielectric material to a dispersionless negative permittivity material based on non-Foster elements. First, the underlying physics of this approach is explored. We consider an initial monochromatic plane wave traveling within an unbounded, spatially homogeneous, isotropic medium having a time-dependent relative permittivity ε(t). A temporal boundary is then induced by rapidly changing ε(t) in a single step from a positive to a dispersionless negative value. It is analytically shown that the phase of the initial monochromatic electromagnetic wave *freezes* in space. While the material is held in this state and the total wave is spatially *frozen*, the amplitudes of the FW and the BW waves excited at the temporal boundary exponentially grows and shrinks in time, respectively, enabling a total wave within the medium (FW+BW wave) to also grow exponentially. In this scenario, the originally monochromatic nature of the wave is transformed into a signal that is no longer periodic in time (while it is still periodic in space) but which has an exponential variation in time, and thus it has no single frequency. To achieve non-dispersive negative permittivity (as required to keep the permittivity negative in time), one must consider non-Foster elements and therefore provide the material with an external energy source. After sufficient time, the permittivity is rapidly rendered back to a positive value greater than unity. At this point, the waves are thawed and resume propagation in both directions. If the new positive permittivity is different from the original positive value, frequency conversion will also occur. Therefore, light amplification and frequency conversion can be achieved with a properly designed temporal function of the permittivity with a negative value over a limited duration of time, forced by an external energy source (as required for non-Foster structures). In addition, we numerically demonstrate a case of a two-dimensional (2D) dipole radiation in such a non-Foster time-varying medium. As an example of the proposed mechanism, a parallel-plate waveguide loaded with switchable parallel thin time-dependent dielectrics emulating capacitors with non-Foster negative capacitances is considered and numerically evaluated demonstrating the ability to achieve the above features.

## *Holding and amplifying waves: rapid change of permittivity from a positive to a negative value*

To begin with, let us discuss the differences and similarities between single-interface spatial and temporal boundaries and between traditional (i.e., passive) and non-Foster (i.e., active with external source) media by examining four cases (see Fig. 1). In all cases, and in the rest of this manuscript, we assume the relative permeability of all materials involved to be that of free space (i.e. $\mu_1 = \mu_2 = 1$), and all permittivity values are real-valued. Our analyses always begin with the traditional incident field, a propagating monochromatic continuous wave (CW) in medium 1, defined as $E_1 = \hat{y} e^{i[\omega_1 t - k_1 x]}$, with $k_1 = \omega_1/v_1$ as the wave number in medium 1, $\omega_1$ is the angular frequency, $v_1 = c/\sqrt{\varepsilon_1}$ is the phase velocity, $c$ is the speed of light in vacuum, and $\varepsilon_1 \geq 1$ is the relative permittivity. Fig. 1a shows the conventional spatial interface between two semi-infinite media defined such that $\varepsilon(x) = \varepsilon_1[x < x_1] + \varepsilon_2[x > x_1]$ where $\varepsilon_2 \geq \varepsilon_1$, defined using Iverson notation. It is well known that upon encountering the spatial interface at $x = x_1$, a reflected and transmitted wave is produced. The transmitted wave number is "modified" to $k_2 = \omega_1/v_2$, with $v_2 = c/\sqrt{\varepsilon_2}$, but the frequencies remain the same (i.e. $\omega_2 = \omega_1$).

Let us now consider the temporal analogue[6,16–18] of such a spatial boundary wherein the medium is spatially unbounded and defined with a time-dependent permittivity $\varepsilon(t) = \varepsilon_1[t < t_1] + \varepsilon_2[t > t_1]$. This scenario is schematically shown in Fig. 1c. When the permittivity of the medium rapidly changes at $t = t_1$ with a rise time much smaller than the period of the signal T, a temporal boundary is introduced[13,16–20,24,25]. Analogous to the spatial boundary, this produces a set of two waves: one traveling forward (FW) (i.e., time-refraction) and one backwards (BW) (i.e., time-reflection). However, in the case of the temporal boundary, the wave number remains unchanged ($k_1 = k_2 = k$) but the frequency of both FW and BW waves is changed from $\omega_1$ to $\omega_2 = (v_2/v_1)\omega_1$ [16,17]. These results are well known[26].

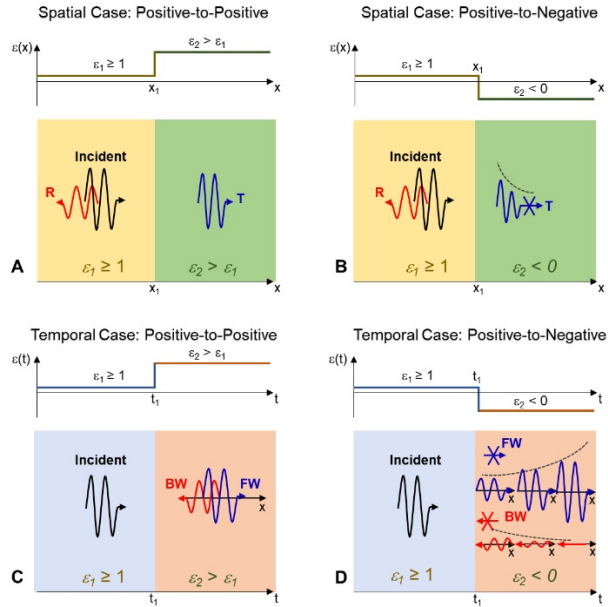

**Figure 1| Schematic representation of analogy between the spatial and temporal boundaries. a,** spatial boundary between two semi-infinite media with two different positive relative permittivities. **b,** spatial boundary between two semi-infinite media with two different oppositely-signed relative permittivities (one positive $\varepsilon_1 \geq 1$ and one negative real value $\varepsilon_2 < 0$). **c,** temporal boundary in a spatially unbounded medium, in which the relative permittivity is temporally changed from a positive value to another positive value. **d,** similar to c, but with relative permittivity is being changed from a positive value to a negative real value (possible only with non-Foster concept) (one positive $\varepsilon_1 \geq 1$ and one negative $\varepsilon_2 < 0$). Note that the sketches of FW and BW waves in c and d represent the field at a single time in the whole space.

Returning to a spatial boundary, consider now the case in which the permittivity of medium 2 is characterized by a negative real value, i.e. $\varepsilon_2 < 0$ (Fig. 1b). In this case, it is well known that the wave number of the transmitted wave in medium 2 becomes imaginary, generating a spatially evanescently decaying field such that there is no propagation. The frequencies of the transmitted, reflected and incident waves remain the same.

However, what happens when the medium on the other side of the *temporal boundary*, $\varepsilon_2$, is negative? See Fig. 1d for a schematic representation of this case. It is expected that, similar to the positive–to–positive temporal interface, a set of FW and BW waves will be generated wherein the wave numbers will be preserved. However, what will happen to the new frequency of these two new waves and how will their propagation be? In this scenario, the frequency is also changed but now from a real $\omega_l$ to an imaginary value (given that $\omega_2 = (v_2/v_1)\omega_1 = (\sqrt{\varepsilon_1}/\sqrt{\varepsilon_2})\omega_1$) leading to an exponential variation in time for the FW and BW waves, with one growing and one decaying in time. In this realm, the total electric ($E_{2\_t1} = E_{FW1} + E_{BW1}$) and magnetic ($H_{2\_t1} = H_{FW1} - H_{BW1}$) fields after inducing the positive–to–negative temporal boundary can be calculated considering the conservation of vectors **D** and **B** at the temporal interface as the known boundary conditions[16] (see Supplementary Materials section 1 and 2 for the detailed formulation):

$$E_{2\_t1} = \hat{y}\left[-\frac{1}{2}\frac{\varepsilon_1}{|\varepsilon_2|}\cos(k_1 x - \omega_1 t_1) - \frac{1}{2}\frac{\sqrt{\varepsilon_1}}{\sqrt{|\varepsilon_2|}}\sin(k_1 x - \omega_1 t_1)\right]e^{\left[\frac{\sqrt{\varepsilon_1}}{\sqrt{|\varepsilon_2|}}\omega_1(t-t_1)\right]}$$
$$+\hat{y}\left[-\frac{1}{2}\frac{\varepsilon_1}{|\varepsilon_2|}\cos(k_1 x - \omega_1 t_1) + \frac{1}{2}\frac{\sqrt{\varepsilon_1}}{\sqrt{|\varepsilon_2|}}\sin(k_1 x - \omega_1 t_1)\right]e^{-\left[\frac{\sqrt{\varepsilon_1}}{\sqrt{|\varepsilon_2|}}\omega_1(t-t_1)\right]}$$

(1a)

$$H_{2\_t1} = \hat{z}\left[\frac{1}{2}\frac{\sqrt{\varepsilon_1 \varepsilon_0}}{\sqrt{\mu_1 \mu_0}}\cos(k_1 x - \omega_1 t_1) - \frac{1}{2}\frac{\sqrt{\varepsilon_0}}{\sqrt{\mu_1 \mu_0}}\frac{\varepsilon_1}{\sqrt{|\varepsilon_2|}}\sin(k_1 x - \omega_1 t_1)\right]e^{\left[\frac{\sqrt{\varepsilon_1}}{\sqrt{|\varepsilon_2|}}\omega_1(t-t_1)\right]}$$
$$+\hat{z}\left[\frac{1}{2}\frac{\sqrt{\varepsilon_1 \varepsilon_0}}{\sqrt{\mu_1 \mu_0}}\cos(k_1 x - \omega_1 t_1) + \frac{1}{2}\frac{\sqrt{\varepsilon_0}}{\sqrt{\mu_1 \mu_0}}\frac{\varepsilon_1}{\sqrt{|\varepsilon_2|}}\sin(k_1 x - \omega_1 t_1)\right]e^{-\left[\frac{\sqrt{\varepsilon_1}}{\sqrt{|\varepsilon_2|}}\omega_1(t-t_1)\right]}$$

(1b)

where $\mu_0$ and $\varepsilon_0$ are the permeability and permittivity of free-space, respectively. The first and second terms of each equation correspond to the FW and BW waves, respectively, although strictly speaking, they are no longer waves, as they are not propagating. Nevertheless, we continue to use the terminology FW and BW for simplicity. The choice of the sign in front of the $i\sqrt{|\varepsilon_2|}$ is justified as we have independently derived the relevant expressions entirely in the time domain using the Laplace transform, as shown in the Supplementary Materials section 8. One immediately notices, from Eq. 1, that the electromagnetic wave is no longer sinusoidally periodic in time (while it is still periodic in space). Rather, the two fields (FW and BW waves) exponentially grow/decay in time and are *frozen* in space. In other words, the electromagnetic wave changes its nature from being monochromatic to one that varies purely exponentially with time with the total electric field $E_{2\_t1}$ not actually moving in the unbounded medium for $t > t_1$. However, as we show later, this field distribution does have a non-zero net time-average Poynting vector, which guarantees the conservation of momentum. This performance is schematically shown in Fig. 1d.

Since the electromagnetic field described by the total electric field in Eq. 1 for $t>t_1$ is no longer monochromatic (and instead it is changing with time purely exponentially) while we desire to keep the value of permittivity negative in this time period, we conclude that this medium is required to be non-dispersive with a constant permittivity forced to be kept negative in this time period. This is unphysical for passive media according to the Foster Reactance theorem[27]. (In the Supplementary Materials section 10, we show an example concerning the impossibility of such negative permittivity for a passive medium with the Drude dispersion.) So far, we have not made any assumption regarding the passive or active nature of the time-dependent metamaterial. However, at this point, it is important to ask: is it indeed possible to have such exponential growth of $E_{2\_t1}$ in a real passive physical system? Interestingly a similar exponential growth has been analytically investigated in a medium with non-dispersive negative permittivity in the time-harmonic scenario[28]. So, what is special about non-dispersive negative permittivity materials? As is known, the basic energy dispersion constraint together with causality requirements forces a passive material to have dispersive parameters satisfying (for the time convention $e^{i\omega t}$) $\{\partial[\omega\varepsilon(\omega)]/\partial\omega\} > \varepsilon_0$ and $\{\partial[\omega\mu(\omega)]/\partial\omega\} > \mu_0$[29]. What these two expressions tell us is that the overall reactive electromagnetic energy stored in any passive lossless/low-loss material is always greater than the energy stored in vacuum. So, for passive media, negative permittivity must be dispersive and cannot be constant over a range of frequencies. Hence, it is straightforward to conclude that, to achieve a non-dispersive negative ε, an active medium with an external source of energy is needed. Interestingly, similar expressions apply to any reactive element in the field of circuit theory. In this realm, the Foster reactance theorem requires that $\{\partial[X(\omega)]/\partial\omega\} > 0$ and $\{\partial[Y(\omega)]/\partial\omega\} > 0$ (with X and Y as the reactance and susceptance, respectively)[27]. This connection between the Foster reactance theorem and the constraints for energy dispersion in passive materials has been exploited to design non-dispersive metamaterials[30–32] such as broadband epsilon-near-zero (ENZ) response[33] by using non-Foster (i.e., active) elements[34,35]. Based on this information, our proposed *frozen* electromagnetic waves mechanism mentioned above may become possible using time-dependent platforms with non-Foster structures with an external source of energy.

## Results for epsilon-positive-to-negative temporal boundary: step function of ε(t)

Let us now evaluate the response of the proposed electromagnetic wave holding phenomenon described by Eq. 1 with some quantitative values. Let us assume that the relative permittivity of the entire medium where the plane wave is traveling is rapidly changed from $\varepsilon_1 = 1$ to $\varepsilon_2 = -20$ at $t_1 = 37.1T$, where T is the period of the original monochromatic wave, and is kept at this value for $t > t_1$.[*] This time-dependent ε(t) is shown in Fig. 2. With this configuration, the analytical results of the total electric and magnetic fields as a function of space and time are shown in Fig. 2d,g, respectively. From these results, we can see how the electromagnetic wave propagates along the positive *x*-axis for $t < t_1$ while it is effectively frozen and exponentially growing once the temporal boundary is introduced (for $t > t_1$). This performance can be also corroborated by looking at the instantaneous Poynting vector (Fig. 2j) as a function of space and time. As observed, it is positive for $t < t_1$ meaning that the wave is propagating, whereas it is spatially frozen and exponentially growing in time for $t > t_1$. (As we show later, when $t > t_1$ the Poynting vector is not strictly zero but instead has a non-zero net value.)

To further investigate these theoretical results, the analytically-evaluated values of the electric and magnetic field distributions along with the instantaneous Poynting vector at $t = 37T$ (before $t_1$) and $t = 39T$ (after $t_1$) are shown in the second (Fig. 2e,h,k) and third columns (Fig. 2f,i,l) of Fig. 2, respectively. These two times correspond to an instant before and after the temporal boundary is introduced at $t_1 = 37.1T$, respectively. As observed, for $t = 37T$ (Fig. 2e,h,k) the electric and magnetic field distributions correspond to those of a single propagating wave along the positive *x*-axis where the electric and magnetic fields are in-phase meaning that the instantaneous Poynting vector is positive, as described before. However, at $t = 39T$ (Fig. 2f,i,l), since the ε has been changed to a negative value and has been kept at this value (using the notion of non-Foster), the electric and magnetic fields are not spatially in synch, meaning that where the E is maximum, the H is almost zero, and vice versa. In fact, as shown in our detailed calculation in the Supplementary Materials section 4, as we change the permittivity from a positive to a negative value at $t = t_1$ the phase difference between the E and H is 180 degrees at $t_{1+}$ (see Fig. 2c), but as time goes on this phase difference approaches 90 degrees (but it never gets to 90 degrees exactly)[†], while the magnitude of E and H grows exponentially.

---

[*] The issue of transitioning through $\varepsilon_2 = 0$ is discussed in detail in Section 9 of the Supplementary Materials.
[†] Although for $t > t_1$ we no longer have a monochromatic signal (but instead we have exponentially growing and decaying fields), nevertheless we still use the notion of phase difference between the E and H to show the spatial separation between their maxima and their

The combination of these two asymptotic behaviors provides a non-zero net Poynting vector for t > $t_1$, which as we will show later, guarantees the conservation of the momentum. For the sake of completeness, the analytical results of the electric and magnetic field distributions for the FW and BW at different times before and after t = $t_1$ are shown in the Supplementary Materials section 5 Fig. S2. From those results it is shown how an exponential growth and decay is experienced by the FW and BW waves, respectively, while they are spatially almost *frozen*. It is worth mentioning that in the work of Lyubarov et al.[23], they achieve amplification of light by periodically modulating the permittivity of the medium in time while the light still propagates, whereas here in our approach we change the permittivity only once in a *single* step in time, and the wave is also essentially frozen in space.

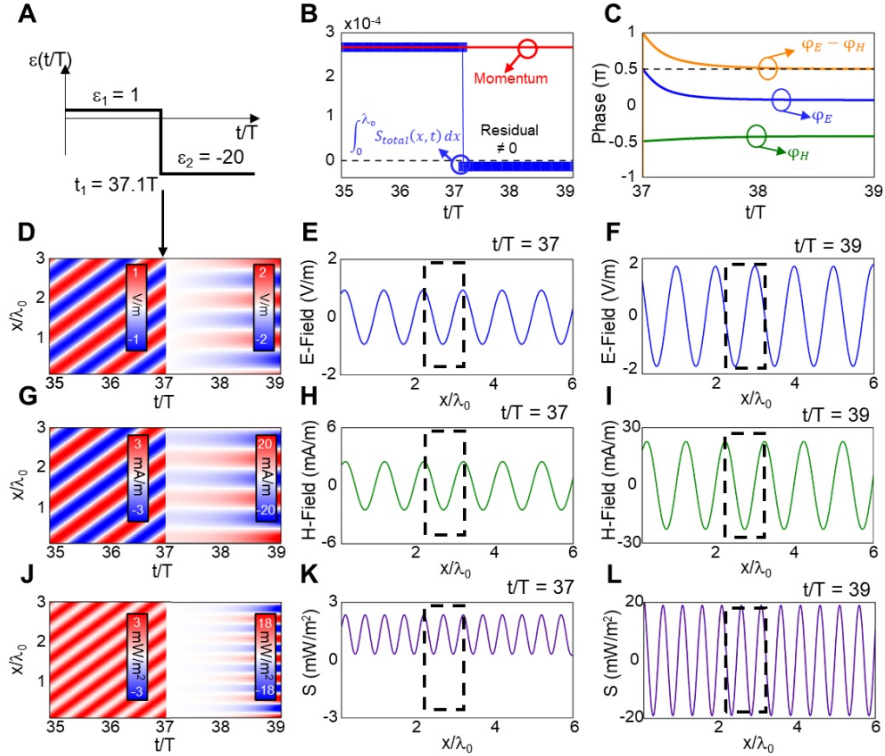

**Figure 2| Analytical results for epsilon-positive -to- negative temporal boundary: step function of ε(t). a,d,g,j** Time-dependent permittivity, electric field, magnetic field, and instantaneous Poynting vector distributions as a function of space (x) and time (t) considering that relative ε of the entire medium is changed from a positive $\varepsilon_1$ = 1 to a negative real value $\varepsilon_2$ = -20 at $t_1$ = 37.1T and is kept at this value (based on non-Foster concept) for t > $t_1$. (Note different ranges of scale bars are used in the two different temporal ranges to show details of the distributions.) **b,** Space average of the instantaneous Poynting vector (blue) over one-wavelength distance ($\int_0^{\lambda_0} S_{total}(x,t)dx$) along with the momentum (red) as a function of time. **c,** Phase of the E (blue) and H fields (green) along with their difference (yellow) for times t > $t_1$ calculated using the theory shown in the Supplementary Materials section 4. **e,f** Electric field, **h,I** magnetic field and **k,l** instantaneous Poynting vector distributions as a function of space (x) extracted from **d,g,j** at a time t = 37T (second column) and t = 39T (third column) corresponding to times before and after the change of permittivity from a positive to a negative value, respectively.

---

zero values, i.e., by approaching 90 degree phase difference here we mean at locations where E is maximum, H is approaching toward its zero (but it will never get there exactly) and vice versa.

### *Results for epsilon-positive-to-negative-to-positive temporal slab: square function of ε(t)*

In the previous section, we discussed the case where ε was rapidly changed from a positive to a negative value and is kept at this value using the concept of the non-Foster phenomenon. However, what would happen if ε is then changed back to a positive value after being negative? This section addresses this question by considering a time dependent ε(t) defined by two single steps (similar to a square function in time, i.e., two consecutive temporal boundaries) i.e., a temporal slab[24]. We consider again a monochromatic plane wave traveling in an unbounded medium with the time-dependent ε as shown in Fig. 3 (the general analytical solution for this scenario can be found in the Supplementary Materials section 1).

We first examine the case where $\varepsilon_1 = 1$ and $\varepsilon_3 > \varepsilon_2 > \varepsilon_1$ meaning that all ε are positive (see Fig. 3a). Here, ε is changed from $\varepsilon_1 = 1$ to $\varepsilon_2 = 5$ at $t_1 = 37T$ and then to $\varepsilon_3 = 10$ at $t_2 = 40T$. With this configuration, the ε of the whole medium is kept as $\varepsilon_2$ for a duration of $t_2 - t_1 = 3T$. The plots of the analytical results of the electric field distribution as a function of space and time for this setup are shown in Fig. 3a. As observed, a set of $FW_1$ and $BW_1$ waves are produced at the first temporal boundary ($t = t_1$) traveling with a new frequency $\omega_2 = (\sqrt{\varepsilon_1}/\sqrt{\varepsilon_2})\omega_1$, as mentioned before. Then, these two waves are again split when the second temporal change of ε is introduced at $t_2$. These four waves (two traveling FW and two traveling BW) propagate now with a single frequency $\omega_3 = (\sqrt{\varepsilon_2}/\sqrt{\varepsilon_3})\omega_2 = (\sqrt{\varepsilon_1}/\sqrt{\varepsilon_3})\omega_1$, as expected. The results of the electric field distribution at different times $t < t_1$, $t_1 < t < t_2$, and $t > t_2$ are shown in Supplementary Materials section 6 (Fig. S3b,c,d), respectively. From these results, it is clear how the frequency of the wave is modified at each temporal boundary but the wave number *k*, and hence the wavelength, is kept the same throughout the process. For the sake of completeness, and to better appreciate the change of frequency after inducing each temporal boundary at $t_1$ and $t_2$, the electric field distribution was calculated at two different locations ($x/\lambda_0 = 2$ and $x/\lambda_0 = 2.25$) and the results are shown in Fig. 3b where it is seen how the period of the wave is different within the time regions $t < t_1$, $t_1 < t < t_2$ and $t > t_2$.

Let us now evaluate the case where ε is brought down to a negative value as shown in Fig. 3d. For this case, ε is rapidly changed from $\varepsilon_1 = 1$ to $\varepsilon_2 = -15$ at $t_1 = 37T$ and it is then changed to a positive value $\varepsilon_3 = 10$ ($\varepsilon_3 \neq \varepsilon_1$) at $t_2 = 40T$. Here, the ε of the entire medium is kept negative (using the of non-Foster scenario) for a duration $t_2 - t_1 = 3T$. The results of the electric field distribution as a function of space and time for this time-dependent ε are shown in Fig. 3d. We see that the electromagnetic wave propagates along the positive *x* for t < $t_1$. Then, the wave is essentially *frozen* for the time interval $t_1 < t < t_2$ and its amplitude exponentially grows (with the FW and BW waves exponentially growing and decaying, as shown in the Supplementary Materials Fig. S3 and Fig. S4) (while it still have a non-zero net Poynting vector, see Fig. 3f), in agreement with the results discussed in the previous section for a step function of ε.

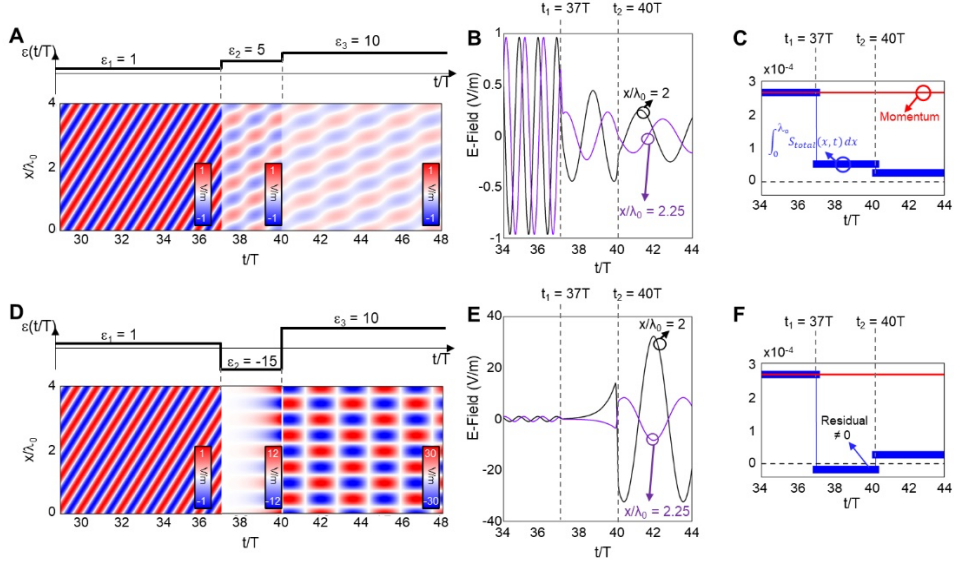

**Figure 3| Analytical results for epsilon-positive-to-negative-to-positive temporal slab: square function of ε(t). a,** Electric field distribution as a function of space (x) and time (t) when relative ε is changed from $\varepsilon_1 = 1$ to a positive real value $\varepsilon_2 = 5$ at $t_1 = 37T$ and it is then changed to $\varepsilon_3 = 10$ at $t_2 = 40T$. **b,** Electric field distribution as a function of time (t) at two spatial locations $x = 2\lambda_0$ (black) and $x = 2.25\lambda_0$ (purple) extracted from panel a. **c,** Space average of the instantaneous Poynting vector over one-wavelength distance ($\int_0^{\lambda_0} S_{total}(x,t)dx$, blue) along with the momentum (red) as a function of time calculated from panel a. **d,** Electric field distribution as a function of space (x) and time (t) when relative ε is changed from $\varepsilon_1 = 1$ to a real negative value $\varepsilon_2 = -15$ (based on the notion of non-Foster) at $t_1 = 37T$ and then to $\varepsilon_3 = 10$ at $t_2 = 40T$. (Note in panel **d** three different ranges of scale bars are used in the three different temporal ranges to show details of the distributions.) **e,** Electric field distribution at $x = 2\lambda_0$ (black) and $x = 2.25\lambda_0$ (purple) extracted from panel d. **f,** same as **c** but for the case discussed in **d**.

Now, at $t > t_2$, the permittivity of the medium is changed back to a positive value $\varepsilon_3$. Remarkably, this causes the wave inside the medium to *thaw*, allowing it to propagate again. However, we note that for $t > t_2$, in addition to the frequency change from $\omega_1$ to $\omega_3 = \left(\sqrt{\varepsilon_1}/\sqrt{\varepsilon_3}\right)\omega_1$, two interesting things happen: i) the wave behaves *almost* like a "standing wave" since the FW and BW waves have approximately (but not exactly) the same amplitude and ii) the amplitude of the electromagnetic wave has been increased up to ~30 times compared to its value for $t < t_1$. It is important to highlight the fact that for $t > t_2$, we do not have a perfect standing wave, because strictly speaking the amplitudes of the FW and the BW waves, albeit very high, are not exactly the same. In fact, we show analytically that there is a net non-zero Poynting vector value in the direction of +x, which guarantees the conservation of momentum. To better observe these results, the electric field distribution at times $t < t_1$, $t_1 < t < t_2$, and $t > t_2$ are also shown in Supplementary Materials section 6 (Fig. S3f,g,h, respectively). For the sake of completeness, the evaluated electric field at the two single locations $x/\lambda_0 = 2$ and $x/\lambda_0 = 2.25$ are also shown in Fig. 3e. Additionally, the instantaneous Poynting vector for the total, FW and BW waves as a function of space and time are shown in the Supplementary Materials Fig. S5. By comparing the results for the negative square function (Fig. 3d-f) with those of the positive square function (Fig. 3a-c) we can see how the phenomena of holding and amplifying the wave and the frequency conversion can be achieved with the proposed temporal slab involving non-dispersive negative permittivity (based on the non-Foster notion).

## *Effect of time interval $t_2 - t_1$ on frozen waves*

Since the frozen waves in the time interval $t_2 - t_1$ experience exponential growth and decay, one may ask: is there any connection between this time interval and the amplitude of the FW and BW waves in $t > t_2$? To answer these questions, we first consider the positive square function of $\varepsilon$ as shown in Fig. 4a. For the sake of simplicity, here $\varepsilon$ is changed from a positive $\varepsilon_1 \geq 1$ to another positive value $\varepsilon_2 > \varepsilon_1$ at $t = t_1$, it is kept to this value for a time $t_2 - t_1$ and then is returned to $\varepsilon_3 = \varepsilon_1 \geq 1$ at $t = t_2$. As discussed in the previous section and as schematically shown in Fig. 4a, a set of waves (FW$_1$, BW$_1$) are produced at the first temporal boundary ($t = t_1$) with coefficients $T_1$ and $R_1$, respectively, which can be easily calculated by applying the temporal continuity of **D** and **B** at $t = t_1$, i.e., $\mathbf{D}_{t1-\delta} = \mathbf{D}_{t1+\delta}$ and $\mathbf{B}_{t1-\delta} = \mathbf{B}_{t1+\delta}$ in the limit when $\delta \to 0$ (see Supplementary Materials Section 1 for full details). Then these two waves are split into a second set (FW$_2$, BW$_2$) at $t = t_2$ with coefficients $T_2$ and $R_2$ calculated following the same process. According to Ref [18], these coefficients $T_1$, $R_1$, $T_2$ and $R_2$ are then defined as $(T_1, R_1) = 0.5[(\varepsilon_1/\varepsilon_2) \pm (\sqrt{\varepsilon_1}/\sqrt{\varepsilon_2})]$ and $(T_2, R_2) = 0.5[(\varepsilon_2/\varepsilon_3) \pm (\sqrt{\varepsilon_2}/\sqrt{\varepsilon_3})]$. The propagating nature of the waves during this whole process has been recently exploited to achieve time reversal for water waves[24] and also antireflection temporal coatings for electromagnetic waves[36] by properly engineering the time interval $t_2 - t_1$. Now, for the negative square function (Fig. 4b) we have seen before in Fig. 3d,e how spatially frozen FW$_1$ and BW$_1$ waves are produced at $t = t_1$ when $\varepsilon$ is changed to a negative value. These frozen waves grow/decay exponentially until they are split into a FW$_2$ and a BW$_2$ at t

= $t_2$ when the permittivity is changed to a positive value again. Considering the scenarios from Fig. 4a,b, the general solution for the total electric field ($E_{3\_t2}$) for a square function of permittivity can be defined as follows (see Supplementary Materials Section 3 for the full formulation along with the complete solution for the magnetic field):

$$E_{3\_t2} = \left\langle \begin{array}{l} \frac{1}{4}\left(\frac{\varepsilon_1}{\varepsilon_3} + \frac{\sqrt{\varepsilon_1}}{\sqrt{\varepsilon_3}}\right) \cos\left[k_1 x - \frac{\sqrt{\varepsilon_1}}{\sqrt{\varepsilon_3}}\omega_1(t-t_2) - \omega_1 t_1\right] \left\{ e^{\left[\frac{\sqrt{\varepsilon_1}}{\sqrt{|\varepsilon_2|}}\omega_1(t_2-t_1)\right]} + e^{-\left[\frac{\sqrt{\varepsilon_1}}{\sqrt{|\varepsilon_2|}}\omega_1(t_2-t_1)\right]} \right\} \\ + \frac{1}{4}\left(-\frac{\varepsilon_1}{\sqrt{|\varepsilon_2|}\sqrt{\varepsilon_3}} + \frac{\sqrt{\varepsilon_1}\sqrt{|\varepsilon_2|}}{\varepsilon_3}\right) \sin\left[k_1 x - \frac{\sqrt{\varepsilon_1}}{\sqrt{\varepsilon_3}}\omega_1(t-t_2) - \omega_1 t_1\right] \left\{ e^{\left[\frac{\sqrt{\varepsilon_1}}{\sqrt{|\varepsilon_2|}}\omega_1(t_2-t_1)\right]} - e^{-\left[\frac{\sqrt{\varepsilon_1}}{\sqrt{|\varepsilon_2|}}\omega_1(t_2-t_1)\right]} \right\} \\ \frac{1}{4}\left(\frac{\varepsilon_1}{\varepsilon_3} - \frac{\sqrt{\varepsilon_1}}{\sqrt{\varepsilon_3}}\right) \cos\left[k_1 x + \frac{\sqrt{\varepsilon_1}}{\sqrt{\varepsilon_3}}\omega_1(t-t_2) - \omega_1 t_1\right] \left\{ e^{\left[\frac{\sqrt{\varepsilon_1}}{\sqrt{|\varepsilon_2|}}\omega_1(t_2-t_1)\right]} + e^{-\left[\frac{\sqrt{\varepsilon_1}}{\sqrt{|\varepsilon_2|}}\omega_1(t_2-t_1)\right]} \right\} \\ + \frac{1}{4}\left(\frac{\varepsilon_1}{\sqrt{|\varepsilon_2|}\sqrt{\varepsilon_3}} + \frac{\sqrt{\varepsilon_1}\sqrt{|\varepsilon_2|}}{\varepsilon_3}\right) \sin\left[k_1 x + \frac{\sqrt{\varepsilon_1}}{\sqrt{\varepsilon_3}}\omega_1(t-t_2) - \omega_1 t_1\right] \left\{ e^{\left[\frac{\sqrt{\varepsilon_1}}{\sqrt{|\varepsilon_2|}}\omega_1(t_2-t_1)\right]} - e^{-\left[\frac{\sqrt{\varepsilon_1}}{\sqrt{|\varepsilon_2|}}\omega_1(t_2-t_1)\right]} \right\} \end{array} \right\rangle \quad (2)$$

where the first and second terms represent the final FW and BW waves (each of them formed by two sets of waves) for $t > t_2$. From Eq. 2, we can see how the exponential terms inside braces define the exponential decay/grow of the frozen waves within the time interval $t_2 - t_1$. Interestingly, it can be seen how the terms $-[(\sqrt{\varepsilon_1}/\sqrt{|\varepsilon_2|})\omega_1(t_2-t_1)]$ and $[(\sqrt{\varepsilon_1}/\sqrt{|\varepsilon_2|})\omega_1(t_2-t_1)]$ will make the exponentials decay and grow in time, respectively, with the decaying term getting smaller for large enough values of $t_2 - t_1$, as discussed in Eq. 1 and Fig. 1-3. Hence, the new set of waves created at $t = t_2$ from this decaying term will be very small (but not equal to zero). The other set of waves created at $t = t_2$ from the growing term will be large. The combination of these sets of waves leads to the FW and BW waves for $t > t_2$ whose amplitudes are approximately (but not exactly) the same, as can be seen from the results in Fig. 3 and Fig. S3 of the Supplementary Materials.

To quantitatively study this performance, let us consider a square function of $\varepsilon$ where it is changed from $\varepsilon_1 = 1$ to a negative real value ($\varepsilon_2$) at $t = t_1$ and then returned to $\varepsilon_3 = \varepsilon_1 = 1$ at $t = t_2$. The results of the amplitude of the frozen FW$_1$ and BW$_1$ waves created after the first temporal boundary at a time just before $t_2$ ($t = t_2^-$) are shown in Fig. 4c,d, respectively, considering different intervals where $\varepsilon$ is kept negative ($t_2 - t_1$). In these plots, two values of $\varepsilon_2$ are considered ($\varepsilon_2 = -4$ and $\varepsilon_2 = -8$) for comparison. The first aspect we observe from these results is that the exponential growth/decay of the frozen FW$_1$ and BW$_1$ is faster for $\varepsilon_2 = -4$ as compared with the case of $\varepsilon_2 = -8$, as expected from Eq. 1. Moreover, we note that when $\varepsilon_2 = -4$, the amplitude of the BW$_1$ decays to ~0.21 when $t_2 - t_1 = 0.2T$ while it is strongly reduced to a value of ~0.01 when increasing $t_2 - t_1$ to 0.8T.

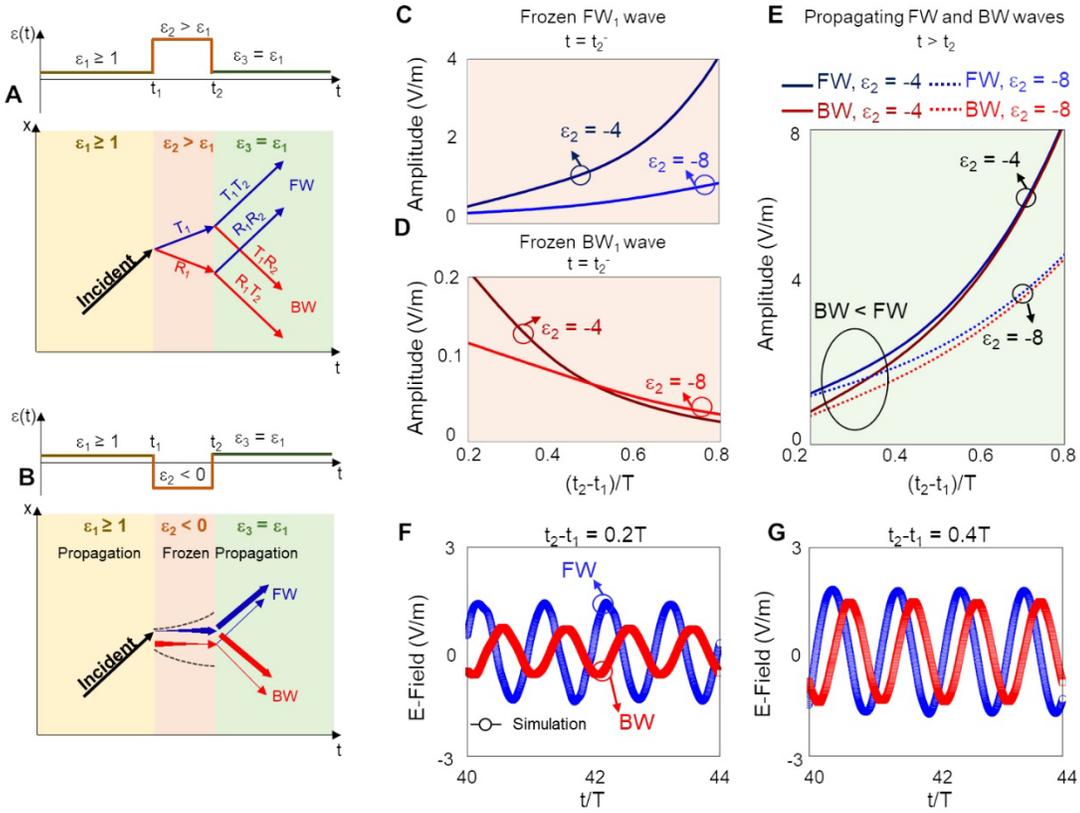

**Figure 4| Effect of the time interval $t_2-t_1$. a,b,** schematic representation of a square function of permittivity considering ($\varepsilon_1 = \varepsilon_3 = 1$, $\varepsilon_2 > 1$) and ($\varepsilon_1 = \varepsilon_3 = 1$, $\varepsilon_2 < 0$), respectively. **c,d,** amplitude of the $FW_1$ and $BW_1$ waves created at the first temporal boundary as a function of the time interval $t_2 - t_1$ where $\varepsilon = \varepsilon_2 < 0$ calculated just before the second temporal boundary ($t_2^-$), respectively. Two values of $\varepsilon_2$ are considered here: $\varepsilon_2 = -4$ (dark blue and dark red) and $\varepsilon_2 = -8$ (light blue and light red). **e,** amplitude of the final FW and BW waves as a function of the time interval $t_2 - t_1$ where $\varepsilon = \varepsilon_2 < 0$ calculated after the second temporal boundary is introduced ($t > t_2$). **f,g,** numerical results of the electric field of FW (blue) and BW (red) waves (calculated at $x/\lambda_0 = 7.5$ and $x/\lambda_0 = -7.5$, respectively) for times $t > t_2$ considering that $\varepsilon_1 = 1$, $\varepsilon_2 = -8$ and $\varepsilon_3 = 1$ using the time intervals $t_2 - t_1 = 0.2T$ and $t_2 - t_1 = 0.4T$ where $\varepsilon = \varepsilon_2 < 0$, respectively.

As explained before, this performance has a direct consequence on the amplitude of the total FW and BW waves created after the second temporal boundary at $t = t_2$ (see Eq. 2). Following the examples shown in Fig. 4c,d, the amplitude of the total FW and BW waves for $t > t_2$ as a function of the time interval $t_2 – t_1$ are shown in Fig 4e considering the same two values of negative $\varepsilon_2$ ($\varepsilon_2 = -4$ and $\varepsilon_2 = -8$). From these results, it is clear how the FW and BW have almost (but not exactly) the same amplitude when using time intervals $t_2 – t_1$ larger than 0.7T whereas this would not be the case when reducing this interval. For the sake of completeness, the simulation results using COMSOL Multiphysics® of the electric field distribution of the FW and BW waves are calculated and shown in Fig. 4f,g, considering a square function of $\varepsilon$ with $\varepsilon_1 = \varepsilon_3 = 1$ and $\varepsilon_2 = -8$ using $t_2 - t_1 = 0.2T$ and $t_2 - t_1 = 0.4T$, respectively. As observed, the amplitudes of the waves are different/similar (but not identical) when using $t_2 - t_1 = 0.2T$ and $t_2 - t_1 = 0.4T$, respectively, in agreement with the discussion from Fig. 4e. For completeness, additional numerical simulations and analytical calculations showing different cases of square functions of permittivity where wave amplification and frequency conversion is achieved with our approach are shown in the Supplementary Materials Fig. S6. Moreover, the analytical results of the instantaneous Poynting vector for cases where $\varepsilon_1 = \varepsilon_3 = 1$ and $\varepsilon_2 = -4$ considering different values of $t_2 - t_1$ (namely 0.2T, 0.4T, and 0.6T) are presented in the Supplementary Materials section 7 Fig. S7, showing how for $t > t_2$ and when $t_2 - t_1 = 0.6T$ the FW and BW wave have similar (but not equal) amplitudes, producing an almost (but not exact) perfect standing wave, as explained above, while it is not the case when a reduced $t_2 - t_1 = 0.2T$ is implemented.

## *Electromagnetic momentum consideration*

In our temporal, spatially unbounded, homogeneous non-Foster metastructures, the electromagnetic momentum must be conserved. This is indeed the case here. Considering $\boldsymbol{D} \times \boldsymbol{B}$ as the electromagnetic momentum per unit volume[37], from our analysis (in the Supplementary Materials) it can be seen from Figs. 2 and 3 that this term stays unchanged throughout all three time regions $t < t_1$, $t_1 < t < t_2$, and $t > t_2$. In temporal regions $t < t_1$ and $t > t_2$, this momentum density, which is directly related to the Poynting vector, is clearly noticed because in these two regions we have propagating waves (i.e., in $t < t_1$ we have a single propagating wave, and in $t > t_2$, we have FW and BW waves with similar (but slightly different) amplitudes, thus we have a net non-zero Poynting vector, see Fig. 2b and Fig. 3f.) However, the temporal region $t_1 < t < t_2$, in which we have frozen waves with growing and decaying fields in time, needs further discussion here. In this time interval, we have these two frozen waves each of which carries zero "space-averaged" Poynting vector (by "space-averaged" we mean the average of the instantaneous Poynting vector over a one-wavelength distance). However, when both frozen waves, one growing and one decaying in time, are considered, this combination exhibits a non-zero net space-averaged Poynting vector that guarantees the conservation of the momentum. The presence of this net non-zero Poynting vector in the combined frozen waves can be interpreted as the temporal analogue of the spatial case of exponentially decaying and growing evanescent waves, where each evanescent wave carries no time-average Poynting vector, whereas the presence of both indeed reveals a net time-average Poynting vector tunneling through these two evanescent waves.

## *Temporal metastructure and non-Foster elements*

As described before, owing to the dispersion constraints for passive media, an active temporal metastructure with an external source of energy is needed to change its permittivity from a positive to a negative non-dispersive value and kept at that value for a range of time. In this section, we propose and numerically evaluate the performance of a 2D parallel-plate waveguide loaded with switchable parallel negative non-Foster elements (emulated by thin time-dependent layers where the ε is changed from a positive, larger than one, to a negative value) to show how the electromagnetic wave in this waveguide can be spatially held and amplified after introducing such active non-Foster elements. In this numerical study, a waveguide filled with air ($\varepsilon_1$= 1) is used working at the frequency of $f$ = 1.5GHz ($\lambda_1$ = 200 mm). The two metallic perfect electric conducting (PEC) plates are separated by a distance $h = 0.5\lambda_1$, the total length along the *x*-axis is $6\lambda_1$ and it is considered that the width (along the *z*-axis, out-of-plane) is W = 1m to approximate it as a 2D waveguide. To emulate the negative permittivity, an array of 75 negative non-Foster capacitors are inserted within the waveguide and separated from each other by an air layer ($\varepsilon_1$= 1) of thickness $0.08\lambda_1$. These negative capacitors are emulated using thin dielectrics (thickness $0.01\lambda_1$) having a time-dependent ε that is changed from a positive ($\varepsilon_1 = 1$) to a negative value $\varepsilon_2 = -4$. With this setup, the emulated negative non-Foster elements can be switched on, changing the permittivity of the medium from a positive to a negative value[33]. The negative capacitors are switched on at $t_1$ = 9T and then kept on for $t > t_1$. The signal from the source is switched off at the same time as the first temporal boundary ($t = t_1$) to ensure the evaluation of the metastructure using the signal already present in the waveguide.

With this configuration, the numerical results of the electric field distribution at different time regions are shown in Fig. 5c-e (for completeness zoom-in figures of the electric field distributions are shown as insets on the right-hand side of Fig. 5c-e). As observed, the wave is traveling along the positive *x* direction for t < $t_1$ (Fig. 5c). Once the temporal boundary is induced at $t_1$ = 9T, the wave inside the region with non-Foster capacitors is essentially frozen in space, and subsequently its amplitude grows in time (see Fig. 5d-e). Note how the electric field inside the waveguide changes its sign when inducing the temporal boundary, in agreement with the discussion above. The energy for such growth is supplied by the external source of energy required for non-Foster elements. These results demonstrate how electromagnetic waves can be held and amplified due to the change in the capacitor values.

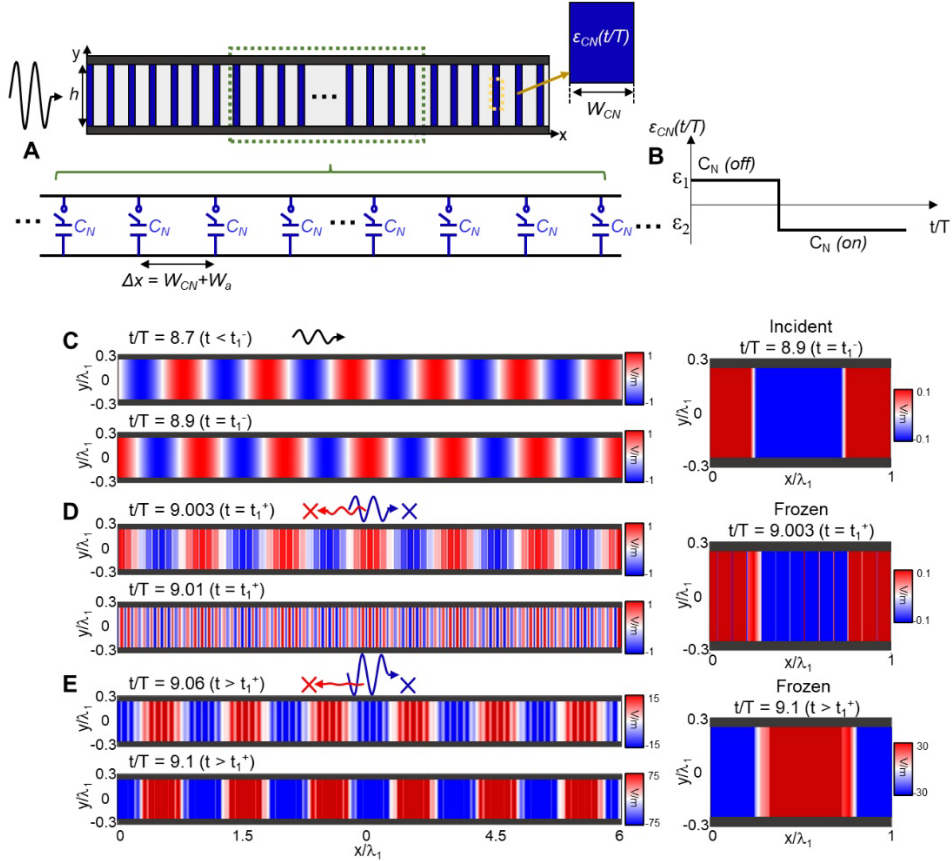

**Figure 5| Proposed idea for a time-dependent parallel-plate waveguides using non-Foster elements. a,** schematic representation of the two-dimensional (2D) parallel-plate waveguide having a height of $h = 0.5\lambda_1$ and a total length along the *x* axis of $6\lambda_1$. A total of 75 non-Foster negative capacitors ($C_N$) are placed within the waveguide separated by a distance $\Delta x = 0.08\lambda_1$. The capacitors are emulated using thin dielectrics of thickness $W_{CN} = 0.01\lambda_1$ having a time dependent ε that is changed from $\varepsilon_1 = 1$ to $\varepsilon_2 = -4$ at $t = t_1 = 9T$. The equivalent circuit of the parallel-plate waveguide is shown in the bottom of panel a. **b,** schematic representation of the change of ε as a function of time. **c,d,e,** numerical results of the electric field distribution at times before and at the temporal boundary ($t < t_1^-$ and $t = t_1^-$), just after the temporal boundary when the capacitors are just activated ($t = t_1^+$), and longer times after the temporal boundary ($t > t_1^+$), respectively. The insets on the right-hand side of panels c-e represent zoom-in figures of the electric field distribution at different times.

## *Dipole radiation in temporal non-Foster metastructures*

As a final study, here we demonstrate the performance of a two-dimensional (2D) dipole immersed within the proposed time-varying non-Foster platform. The schematic representation is shown in Fig. 6a. For times $t < t_1$ the electric 2D dipole is radiating within the unbounded medium of $\varepsilon_1 = 1$. At $t = t_1$, the dipole is switched off and the ε is rapidly changed to $\varepsilon_2 < 0$, meaning that the electromagnetic signal already present in the medium will be held in space and its magnitude will increase exponentially. The numerical results (using COMSOL Multiphysics ®) of this scenario considering that $\varepsilon_1 = 1$ for $t < t_1$ and it is changed to $\varepsilon_2 = $

-4 at t = $t_1$ = 9.4T are shown in Fig. 6b (where the E field, H field, and power distributions as a function of time are shown, all calculated at $x/\lambda_0 = y/\lambda_0 = 0.75$. To enable the convergence of the results, the source was switched off first, and then 0.7T later ε was rapidly changed (see method section for further details). As observed, for times t < $t_1$ the wave emitted by the 2D dipole propagates within the unbounded medium while it is frozen and exponentially increases for t > $t_1$, in agreement with the discussion presented in the previous sections.

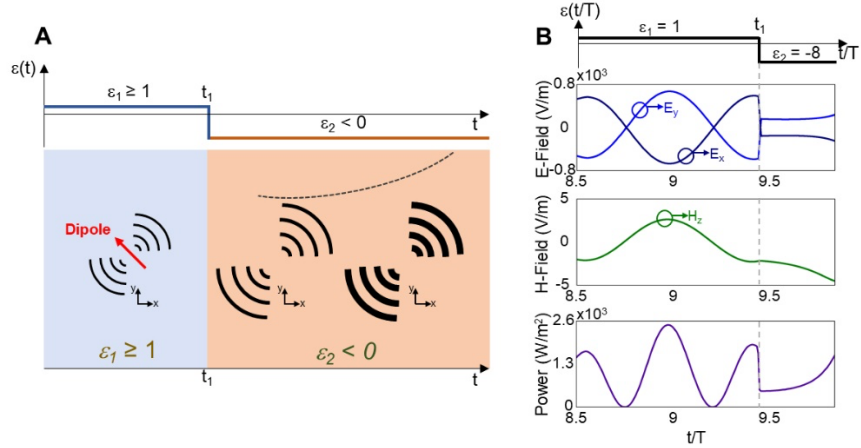

**Figure 6| Immersing a two-dimensional (2D) dipole within the temporal non-Foster platform. a,** Schematic representation showing the dipole radiating for times t < $t_1$ where $\varepsilon_1 \geq 1$ and then the signal is spatially frozen and exponentially increasing its magnitude for times t > $t_1$ where $\varepsilon_2 < 0$. **b,** Numerical results of the in-plane components of the E field ($E_x$, dark blue, and $E_y$, blue), out-of-plane H field (green) and power distribution (purple) calculated at $x/\lambda_0 = y/\lambda_0 = 0.75$. The time-dependent ε function is shown on the top panel from **b.**

In conclusion, we have explored scenarios where temporal boundaries are induced by using temporal metastructures with a rapid change of permittivity from a positive to a negative value, requiring non-Foster structures with external energy. The underlying physics of such temporal structures have been studied here showing how the electromagnetic wave present within such a medium can be forced to effectively "stop", achieving a *frozen* status, and then grow in time, as demonstrated, analytically and numerically, here. Moreover, if the permittivity of the temporal metastructure is returned to a different positive value, propagation is again allowed while also achieving frequency conversion. It has been shown how an active medium with external energy is required to achieve such non-dispersive negative permittivity in a temporal structure, which has a direct relation with the well-known Foster reactance theorem in circuit theory. An example of our proposed theory for such "filling stations for waves" has been studied numerically using a parallel plate waveguide loaded with parallel thin time-dependent media emulating non-Foster negative capacitors demonstrating agreement between the analytical and numerical calculations. This concept may have exciting potential applications in slowing and stopping light combined with amplification, which may lead to new possibilities for coherent sources.

**Methods**

All the numerical simulations were carried out using the time-domain solver of the commercial software COMSOL Multiphysics®. For the configuration shown in Fig. 4-5 along with those results shown in the Supplementary Materials Fig. S6, PEC boundaries were used for the top and bottom boundaries while scattering boundary conditions were implemented at the right boundary to avoid undesirable reflections. The incident field was applied from the left boundary of the simulation box via a scattering boundary condition. For the results shown in Fig. 4 and Fig. S6 a refined Mapped mesh was implemented with a maximum and minimum element size of $0.14\lambda_0$ and $0.15\times10^{-3}\lambda_0$, respectively, with curvature factor of 0.2 and a resolution of narrow regions of 1. The rapid changes of ε from these figures were modeled by implementing an analytical function that combined two built-in single step functions (i.e., emulating square functions ε) having a smooth transition of $7.5\times10^{-6}$ T and one derivative to ensure convergence of results. For the results discussed in Fig. 5, a Mapped mesh was also applied with a maximum and minimum element size of $0.16\lambda_0$ and $0.15\times10^{-3}\lambda_0$, and the same curvature factor and resolution of narrow region values as for the results in Fig. 4. An extra automatic mesh refinement was also implemented. The rapid change of ε was modeled using an analytical single step function having a smooth transition of $1.2\times10^{-5}$ T and one continuous derivative to ensure convergence of results. For the results shown in Fig. 6, an electric dipole was placed at $x/\lambda_0 = y/\lambda_0 = 0$ and embedded within a circular geometry (modeling the unbounded medium) with a radius of $3\lambda_0$. Scattering boundary conditions around the circle were implemented to avoid undesirable reflections. The dipole was modeled using +1 and -1 dipole moment along the *x*- and *y*-axes, respectively, to model a 45 degrees rotated dipole. The electric current dipole moment magnitude was modeled using a sinusoidal signal being switched off at t/T = 8.7 using a smooth transition of ∼1.49T and one continuous derivative to aid in the convergence of the simulation. The ε of the whole medium was rapidly from $\varepsilon_1 = 1$ to $\varepsilon_2 = -8$ at $t_1 = 9.4$T using an analytical step function with a smoothing of $2.9\times10^{-4}$ T and one continuous derivative.

## Acknowledgements

The authors would like to acknowledge the partial support from the Vannevar Bush Faculty Fellowship program sponsored by the Basic Research Office of the Assistant Secretary of Defense for Research and Engineering, funded by the Office of Naval Research through grant N00014-16-1-2029. V.P.-P. acknowledges support from the Newcastle University (Newcastle University Research Fellowship). N.E. also acknowledges partial support from the Simons Foundation/Collaboration on Symmetry-Driven Extreme Wave Phenomena grant # 733684

# Supplementary Materials

## *Holding and amplifying electromagnetic waves with temporal non-Foster metastructures*

*Victor Pacheco-Peña[1], Yasaman Kiasat[2], Diego M. Solís[2,3], Brian Edwards[2] and Nader Engheta[2*]*

[1]School of Mathematics, Statistics and Physics, Newcastle University, Newcastle Upon Tyne, NE1 7RU, UK
[2] Department of Electrical and Systems Engineering, University of Pennsylvania, Philadelphia, PA 19104, USA
[3] Departamento de Tecnología de los Computadores y de las Comunicaciones, University of Extremadura, 10003 Cáceres, Spain
[*]email: engheta@seas.upenn.edu

1. Inducing two temporal boundaries: analytical formulation.

2. Single step function of permittivity $\varepsilon_1 \geq 1$ and $\varepsilon_2 < 0$: analytical formulation.

3. Square function of permittivity $\varepsilon_1 \geq 1$, $\varepsilon_2 < 0$ and $1 \leq \varepsilon_3 \neq \varepsilon_1$: analytical formulation.

4. E and H phase difference when $\varepsilon_2 < 0$.

5. Forward and backward waves results: single step function of permittivity.

6. Forward and backward waves results: square function of permittivity.

7. Instantaneous Poynting vector: effect of $t_2 - t_1$.

8. Laplace transform derivation of electric field after temporal boundary at $t = 0$.

9. Non-Foster structures with loss.

   9.1 Study of Stability.

   9.2 Amplitudes of the waves scattered *at $t=t_1$*.

   9.3 "Unfreezing" the *forward and backward* waves *at $t=t_2$*.

10. Impossibility of rapid temporal variation of permittivity (from positive to negative values) in Drude passive media.

11. Supplementary Movies

12. References

## 1. Inducing two temporal boundaries: analytical formulation

Here we present the analytical formulation for the case where two temporal boundaries are induced. The problem under study is shown in Fig. 1c,d of the main manuscript and it is schematically shown in Fig. S1. It is considered that a monochromatic continuous wave (CW) is traveling in an unbounded medium whose permittivity is time-dependent ε(t). This ε has an initial relative value of $\varepsilon_1$ for $t < t_1$. Then, it is rapidly changed to $\varepsilon_2$ at $t = t_1$ (inducing the first temporal boundary), it is kept at this value for a time interval $t_2 - t_1$ and then it is rapidly changed to $\varepsilon_3$ at $t = t_1$ (where the second temporal boundary is introduced). It is important to highlight that this rapid change of permittivity is not instantaneous, but it is fast enough to possess a transient rise time much less than the period of the wave.

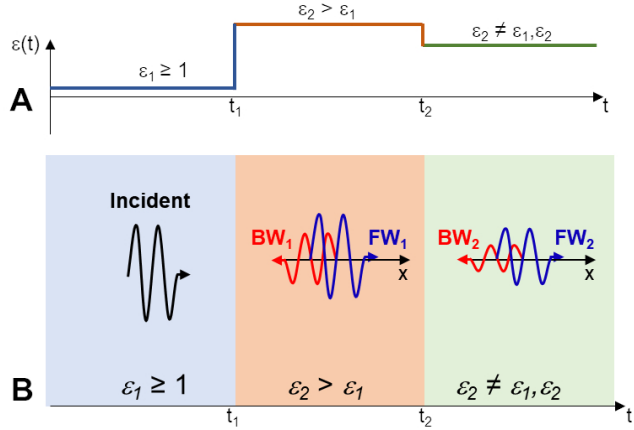

**Figure S1| Schematic representation of the two induced temporal boundaries. a,** Time-dependent permittivity. **b,** Time-dependent permittivity and spatial waves produced at each temporal boundary.

With this setup, it is possible to define the electric field within different temporal windows: $t < t_1$, $t_1 < t < t_2$ and $t > t_2$ where the permittivity of the whole medium is $\varepsilon_1$, $\varepsilon_2$ and $\varepsilon_3$, respectively. Let us first start for the times $t < t_1$. Here, the electric ($E$) and magnetic ($H$) fields of the monochromatic CW traveling in the unbounded medium can be defined as follows:

$$E_1 = \hat{y} e^{i[\omega_1 t - (k_1 x)]} \tag{S.1a}$$

$$H_1 = \hat{z} \frac{\sqrt{\varepsilon_0}}{\sqrt{\mu_0}} \frac{\sqrt{\varepsilon_1}}{\sqrt{\mu_1}} e^{i[\omega_1 t - (k_1 x)]} \tag{S.1b}$$

where $\omega_1$ is the angular frequency, $k_1 = \omega_1/v_1$ is the wave number, $v_1 = c/\sqrt{\mu_1 \varepsilon_1}$ is the phase velocity, $c$ is the speed of light in vacuum and $\varepsilon_1$ and $\mu_1$ are the relative permittivity and permeability, respectively.

Now, at $t = t_1$, the $\varepsilon$ is rapidly changed from $\varepsilon_1$ to $\varepsilon_2$. In this scenario, the first temporal boundary is induced affecting the wave already present in the entire medium. In this context, a forward wave ($E_{2FW\_t1}$) and a backward wave ($E_{2BW\_t1}$) are produced, which are the temporal version of the transmitted and reflected waves in the spatial case for wave propagation between two semi-infinite media with different impedances.

The total electric field ($E_{2\_t1}$) in the whole medium can then be written as follows (we will limit here to write only the $E$-field):

$$E_{2\_t1} = E_{2FW\_t1} + E_{2BW\_t1} \quad (S.2a)$$

$$E_{2\_t1} = T_1 e^{i\omega_1 t_1} e^{i\left[\frac{v_2}{v_1}\omega_1(t-t_1)\right]} e^{-i(k_1 x)} + R_1 e^{i\omega_1 t_1} e^{-i\left[\frac{v_2}{v_1}\omega_1(t-t_1)\right]} e^{-i(k_1 x)} \quad (S.2b)$$

where $T_1$ and $R_1$ are the coefficients of the FW and BW waves, respectively, and $v_2 = c/\sqrt{\mu_2 \varepsilon_2}$ is the phase velocity. The coefficients $T_1$ and $R_1$ can be calculated by applying the conservation of **D** and **B** at $t = t_1$, i.e., $D_{t1-\delta} = D_{t1+\delta}$ and $B_{t1-\delta} = B_{t1+\delta}$ in the limit when $\delta \to 0$. From now on we will consider that only the permittivity is changed in time, i.e., $\mu_1 = \mu_2 = 1$. Based on this, these two coefficients can be defined as follows[1–3]:

$$T_1 = \frac{1}{2}\left(\frac{\varepsilon_1}{\varepsilon_2} + \frac{\sqrt{\varepsilon_1}}{\sqrt{\varepsilon_2}}\right) \quad (S.3a)$$

$$R_1 = \frac{1}{2}\left(\frac{\varepsilon_1}{\varepsilon_2} - \frac{\sqrt{\varepsilon_1}}{\sqrt{\varepsilon_2}}\right) \quad (S.3b)$$

As observed from Eq. S.2 the FW and BW waves created at the induced temporal interface travel with the new frequency ($v_2/v_1$)$\omega_1$ while the wave number $k$ is preserved. These are two important features of temporal boundaries[2] which are different to spatial boundaries where the opposite (conservation of frequency and change of $k$) is expected.

Now, at $t = t_2$, the permittivity is rapidly changed from $\varepsilon_2$ to $\varepsilon_3$. This will introduce a second temporal boundary and the two waves (FW$_1$, BW$_1$) already traveling in the whole medium ($E_{2FW\_t1} + E_{2BW\_t1}$) will be again split into four new waves: two traveling forward and two traveling backward. In this context, the total field in the whole medium ($E_{3\_t2}$) can be then calculated as follows:

$$\begin{aligned}E_{3\_t2} &= T_1 T_2 e^{i\omega_1 t_1} e^{i\left[\frac{v_2}{v_1}\omega_1(t_2-t_1)\right]} e^{i\frac{v_3}{v_1}\omega_1(t-t_2)} e^{-i(k_1 x)} + R_1 R_2 e^{i\omega_1 t_1} e^{-i\left[\frac{v_2}{v_1}\omega_1(t_2-t_1)\right]} e^{i\frac{v_3}{v_1}\omega_1(t-t_2)} e^{-i(k_1 x)} \\ &+ T_1 R_2 e^{i\omega_1 t_1} e^{i\left[\frac{v_2}{v_1}\omega_1(t_2-t_1)\right]} e^{-i\frac{v_3}{v_1}\omega_1(t-t_2)} e^{-i(k_1 x)} + R_1 T_2 e^{i\omega t_1} e^{-i\left[\frac{v_2}{v_1}\omega_1(t_2-t_1)\right]} e^{-i\frac{v_3}{v_1}\omega_1(t-t_2)} e^{-i(k_1 x)}\end{aligned}$$

$$(S.4)$$

with $v_3 = c/\sqrt{\mu_3 \varepsilon_3}$ is the phase velocity and $T_2$ and $R_2$ are the coefficients for the new set of FW and BW waves at $t = t_2$. These coefficients can be calculated following the same process as in Eq. S.3:

$$T_2 = \frac{1}{2}\left(\frac{\varepsilon_2}{\varepsilon_3} + \frac{\sqrt{\varepsilon_2}}{\sqrt{\varepsilon_3}}\right) \quad \text{(S.5a)}$$

$$R_2 = \frac{1}{2}\left(\frac{\varepsilon_2}{\varepsilon_3} - \frac{\sqrt{\varepsilon_2}}{\sqrt{\varepsilon_3}}\right) \quad \text{(S.5b)}$$

The formulation shown in this section represents the solution considering arbitrary values for $\varepsilon_1$, $\varepsilon_2$ and $\varepsilon_3$. In the next sections we provide details for the specific case presented in the main text where $\varepsilon_2 < 0$ and $\varepsilon_1, \varepsilon_3 \geq 1$ in order to *freeze* electromagnetic waves.

## 2. Single step function of permittivity $\varepsilon_1 \geq 1$ and $\varepsilon_2 < 0$: analytical formulation.

Now that we have defined the electric field when inducing two temporal boundaries, let us focus our attention on the total electric field considering the special case when the permittivity is rapidly changed from a positive ($\varepsilon_1 \geq 1$) to a negative value $\varepsilon_2 < 0$ at $t = t_1$ and is kept to this value, i.e., a single step function of permittivity, using an active non-Foster structure with external source of energy.

In this scenario, the total electric field before and after the change of permittivity can be described by Eq. S.1a and Eq. S.2, respectively. After rearranging Eq. S.2, the electric field for $t > t_1$ is as follows:

$$E_{2\_t1} = T_1 e^{i\left[\frac{v_2}{v_1}\omega_1(t-t_1)\right]} e^{-i(k_1 x - \omega_1 t_1)} + R_1 e^{-i\left[\frac{v_2}{v_1}\omega_1(t-t_1)\right]} e^{-i(k_1 x - \omega_1 t_1)} \quad (S.6)$$

Now, we can work with Eq. S.6, considering the case under study where the permittivity is changed from a positive ($\varepsilon_1 \geq 1$) to $\varepsilon_2 < 0$. With this setup, the total field for $t > t_1$ is then defined as:

$$E_{2\_t1} = Re\left\{T_1 e^{\left[\frac{\sqrt{\varepsilon_1}}{\sqrt{|\varepsilon_2|}}\omega_1(t-t_1)\right]} e^{-i(k_1 x - \omega_1 t_1)} + R_1 e^{-\left[\frac{\sqrt{\varepsilon_1}}{\sqrt{|\varepsilon_2|}}\omega_1(t-t_1)\right]} e^{-i(k_1 x - \omega_1 t_1)}\right\} \quad (S.7a)$$

$$E_{2\_t1} = Re\left\{T_1 e^{\left[\frac{\sqrt{\varepsilon_1}}{\sqrt{|\varepsilon_2|}}\omega_1(t-t_1)\right]} [\cos(k_1 x - \omega_1 t_1) - i\sin(k_1 x - \omega_1 t_1)] \right.$$
$$\left. + R_1 e^{-\left[\frac{\sqrt{\varepsilon_1}}{\sqrt{|\varepsilon_2|}}\omega_1(t-t_1)\right]} [\cos(k_1 x - \omega_1 t_1) - i\sin(k_1 x - \omega_1 t_1)]\right\} \quad (S.7b)$$

$$E_{2\_t1} = \left[-\frac{1}{2}\frac{\varepsilon_1}{|\varepsilon_2|}\cos(k_1 x - \omega_1 t_1) - \frac{1}{2}\frac{\sqrt{\varepsilon_1}}{\sqrt{|\varepsilon_2|}}\sin(k_1 x - \omega_1 t_1)\right] e^{\left[\frac{\sqrt{\varepsilon_1}}{\sqrt{|\varepsilon_2|}}\omega_1(t-t_1)\right]}$$
$$+ \left[-\frac{1}{2}\frac{\varepsilon_1}{|\varepsilon_2|}\cos(k_1 x - \omega_1 t_1) + \frac{1}{2}\frac{\sqrt{\varepsilon_1}}{\sqrt{|\varepsilon_2|}}\sin(k_1 x - \omega_1 t_1)\right] e^{-\left[\frac{\sqrt{\varepsilon_1}}{\sqrt{|\varepsilon_2|}}\omega_1(t-t_1)\right]} \quad (S.7c)$$

Note that the first and second terms in Eq. S.7c correspond to the FW and BW waves, respectively, produced after the temporal boundary at $t = t_1$. As observed, these two waves grow/decay exponentially in time while they are spatially *frozen*. The analytical results of the FW and BW waves for the example shown in Fig. 2 of the main manuscript are shown in Fig. S2.

For completeness, we can calculate the magnetic field $H_{2\_t1} = H_{2FW\_t1} - H_{2BW\_t1}$ from Eq. S.7a as follows:

$$H_{2\_t1} = Re\left\{\eta_2 T_1 e^{\left[\frac{\sqrt{\varepsilon_1}}{\sqrt{|\varepsilon_2|}}\omega_1(t-t_1)\right]} e^{-i(k_1 x - \omega_1 t_1)} - \eta_2 R_1 e^{-\left[\frac{\sqrt{\varepsilon_1}}{\sqrt{|\varepsilon_2|}}\omega_1(t-t_1)\right]} e^{-i(k_1 x - \omega_1 t_1)}\right\} \quad (S.7d)$$

with $\eta_2 = i\left(\sqrt{\mu_1\mu_0}/\sqrt{|\varepsilon_2|\varepsilon_0}\right)$. Following the same process as Eqs. S.7a-c, we arrive at the following expression:

$$\begin{aligned}H_{2\_t1} = &\hat{z}\left[\frac{1}{2}\frac{\sqrt{|\varepsilon_2|\varepsilon_0}}{\sqrt{\mu_1\mu_0}}\frac{\sqrt{\varepsilon_1}}{\sqrt{|\varepsilon_2|}}\cos(k_1 x - \omega_1 t_1) - \frac{1}{2}\frac{\sqrt{|\varepsilon_2|\varepsilon_0}}{\sqrt{\mu_1\mu_0}}\frac{\varepsilon_1}{|\varepsilon_2|}\sin(k_1 x - \omega_1 t_1)\right] e^{\left[\frac{\sqrt{\varepsilon_1}}{\sqrt{|\varepsilon_2|}}\omega_1(t-t_1)\right]}\\ +&\hat{z}\left[\frac{1}{2}\frac{\sqrt{|\varepsilon_2|\varepsilon_0}}{\sqrt{\mu_1\mu_0}}\frac{\sqrt{\varepsilon_1}}{\sqrt{|\varepsilon_2|}}\cos(k_1 x - \omega_1 t_1) + \frac{1}{2}\frac{\sqrt{|\varepsilon_2|\varepsilon_0}}{\sqrt{\mu_1\mu_0}}\frac{\varepsilon_1}{|\varepsilon_2|}\sin(k_1 x - \omega_1 t_1)\right] e^{-\left[\frac{\sqrt{\varepsilon_1}}{\sqrt{|\varepsilon_2|}}\omega_1(t-t_1)\right]}\end{aligned} \quad (S.7e)$$

Corresponding to Eq. 1b from the main text.

## 3. Square function of permittivity $\varepsilon_1 \geq 1$, $\varepsilon_2 < 0$ and $1 \leq \varepsilon_3 \neq \varepsilon_1$: analytical formulation.

As detailed in the main text, one of the interesting features of the proposed mechanism to spatially *freeze* electromagnetic wave propagation is that both frequency conversion along with wave amplification can be achieved once the time-dependent permittivity is changed from positive ($\varepsilon_1 \geq 1$) to negative ($\varepsilon_2 < 0$) and then returned to a different positive value ($1 \leq \varepsilon_3 \neq \varepsilon_1$). In this section we provide details on the analytical formulation for such exciting case.

The total electric field when the permittivity is changed from positive to a real negative was defined in Eq. S.7e while the total electric field when the permittivity is returned to a positive ($t > t_2$) is defined by Eq. S.4. After substituting $\varepsilon_2 < 0$ in Eq. S.4, the equation that defines the electric field for $t > t_2$ is as follows:

$$E_{3\_t2} = Re\left\{e^{-i(k_1 x)}e^{i\frac{\sqrt{\varepsilon_1}}{\sqrt{\varepsilon_3}}\omega_1(t-t_2)}e^{i\omega t_1}\left\{T_1 T_2 e^{\left[\frac{\sqrt{\varepsilon_1}}{\sqrt{|\varepsilon_2|}}\omega_1(t_2-t_1)\right]} + R_1 R_2 e^{-\left[\frac{\sqrt{\varepsilon_1}}{\sqrt{|\varepsilon_2|}}\omega_1(t_2-t_1)\right]}\right\}\right.$$
$$\left. + e^{-i(k_1 x)}e^{-i\frac{\sqrt{\varepsilon_1}}{\sqrt{\varepsilon_3}}\omega_1(t-t_2)}e^{i\omega t_1}\left\{T_1 R_2 e^{\left[\frac{\sqrt{\varepsilon_1}}{\sqrt{|\varepsilon_2|}}\omega_1(t_2-t_1)\right]} + R_1 T_2 e^{-\left[\frac{\sqrt{\varepsilon_1}}{\sqrt{|\varepsilon_2|}}\omega_1(t_2-t_1)\right]}\right\}\right\} \quad (S.8a)$$

Note that the first and second term (line) in the expression above correspond to the final FW and BW wave produced after the second temporal boundary, respectively. As observed in Eq. S.8, the total electric field depend on the coefficients $T_1 T_2$, $R_1 R_2$, $T_1 R_2$ and $R_1 T_2$ which can be calculated using Eq. S.3 and Eq. S.5. Based on this, after working with Eq. S.8, the final solution for the total electric field for $t > t_2$ is as follows:

$$E_{3\_t2} = \left\{\frac{1}{4}\left(\frac{\varepsilon_1}{\varepsilon_3} + \frac{\sqrt{\varepsilon_1}}{\sqrt{\varepsilon_3}}\right) cos\left[k_1 x - \frac{\sqrt{\varepsilon_1}}{\sqrt{\varepsilon_3}}\omega_1(t-t_2) - \omega_1 t_1\right]\left\{e^{\left[\frac{\sqrt{\varepsilon_1}}{\sqrt{|\varepsilon_2|}}\omega_1(t_2-t_1)\right]} + e^{-\left[\frac{\sqrt{\varepsilon_1}}{\sqrt{|\varepsilon_2|}}\omega_1(t_2-t_1)\right]}\right\}\right.$$
$$+ \frac{1}{4}\left(-\frac{\varepsilon_1}{\sqrt{|\varepsilon_2|}\sqrt{\varepsilon_3}} + \frac{\sqrt{\varepsilon_1}\sqrt{|\varepsilon_2|}}{\varepsilon_3}\right) sin\left[k_1 x - \frac{\sqrt{\varepsilon_1}}{\sqrt{\varepsilon_3}}\omega_1(t-t_2) - \omega_1 t_1\right]\left\{e^{\left[\frac{\sqrt{\varepsilon_1}}{\sqrt{|\varepsilon_2|}}\omega_1(t_2-t_1)\right]} - e^{-\left[\frac{\sqrt{\varepsilon_1}}{\sqrt{|\varepsilon_2|}}\omega_1(t_2-t_1)\right]}\right\}$$
$$\left\{+ \frac{1}{4}\left(\frac{\varepsilon_1}{\varepsilon_3} - \frac{\sqrt{\varepsilon_1}}{\sqrt{\varepsilon_3}}\right) cos\left[k_1 x + \frac{\sqrt{\varepsilon_1}}{\sqrt{\varepsilon_3}}\omega_1(t-t_2) - \omega_1 t_1\right]\left\{e^{\left[\frac{\sqrt{\varepsilon_1}}{\sqrt{|\varepsilon_2|}}\omega_1(t_2-t_1)\right]} + e^{-\left[\frac{\sqrt{\varepsilon_1}}{\sqrt{|\varepsilon_2|}}\omega_1(t_2-t_1)\right]}\right\}\right.$$
$$\left. + \frac{1}{4}\left(\frac{\varepsilon_1}{\sqrt{|\varepsilon_2|}\sqrt{\varepsilon_3}} + \frac{\sqrt{\varepsilon_1}\sqrt{|\varepsilon_2|}}{\varepsilon_3}\right) sin\left[k_1 x + \frac{\sqrt{\varepsilon_1}}{\sqrt{\varepsilon_3}}\omega_1(t-t_2) - \omega_1 t_1\right]\left\{e^{\left[\frac{\sqrt{\varepsilon_1}}{\sqrt{|\varepsilon_2|}}\omega_1(t_2-t_1)\right]} - e^{-\left[\frac{\sqrt{\varepsilon_1}}{\sqrt{|\varepsilon_2|}}\omega_1(t_2-t_1)\right]}\right\}\right\} \quad (S.8b)$$

Retrieving the expression shown in Eq. 2 from the main text.

For completeness, the magnetic field $H_{2\_t2}$ can be calculated from Eq. S.8a following the same process as in Eq. S.7d:

$$H_{2\_t1} = \begin{aligned} &Re\left\{\eta_3 e^{-i(k_1 x)} e^{i\frac{\sqrt{\varepsilon_1}}{\sqrt{\varepsilon_3}}\omega_1(t-t_2)} e^{i\omega_1 t_1}\left\{T_1 T_2 e^{\left[\frac{\sqrt{\varepsilon_1}}{\sqrt{|\varepsilon_2|}}\omega_1(t_2-t_1)\right]} + R_1 R_2 e^{-\left[\frac{\sqrt{\varepsilon_1}}{\sqrt{|\varepsilon_2|}}\omega_1(t_2-t_1)\right]}\right\}\right.\\ &\left. -\eta_3 e^{-i(k_1 x)} e^{-i\frac{\sqrt{\varepsilon_1}}{\sqrt{\varepsilon_3}}\omega_1(t-t_2)} e^{i\omega t_1}\left\{T_1 R_2 e^{\left[\frac{\sqrt{\varepsilon_1}}{\sqrt{|\varepsilon_2|}}\omega_1(t_2-t_1)\right]} + R_1 T_2 e^{-\left[\frac{\sqrt{\varepsilon_1}}{\sqrt{|\varepsilon_2|}}\omega_1(t_2-t_1)\right]}\right\}\right\} \end{aligned} \quad (S8.c)$$

with $\eta_3 = \left(\sqrt{\mu_1\mu_0}/\sqrt{\varepsilon_3\varepsilon_0}\right)$ and $(\mu_1 = \mu_2 = \mu_3)$. Following the same process as Eqs. S.8a-b, we arrive to the following expression:

$H_{2\_t1} =$

$$\begin{aligned}&\frac{\sqrt{\mu_1\mu_0}}{\sqrt{\varepsilon_3\varepsilon_0}}\left\{\frac{1}{4}\left(\frac{\varepsilon_1}{\varepsilon_3}+\frac{\sqrt{\varepsilon_1}}{\sqrt{\varepsilon_3}}\right)\cos\left[k_1 x - \frac{\sqrt{\varepsilon_1}}{\sqrt{\varepsilon_3}}\omega_1(t-t_2) - \omega_1 t_1\right]\left\{e^{\left[\frac{\sqrt{\varepsilon_1}}{\sqrt{|\varepsilon_2|}}\omega_1(t_2-t_1)\right]} + e^{-\left[\frac{\sqrt{\varepsilon_1}}{\sqrt{|\varepsilon_2|}}\omega_1(t_2-t_1)\right]}\right\}\right.\\ &\left.+\frac{1}{4}\left(-\frac{\varepsilon_1}{\sqrt{|\varepsilon_2|}\sqrt{\varepsilon_3}}+\frac{\sqrt{\varepsilon_1}\sqrt{|\varepsilon_2|}}{\varepsilon_3}\right)\sin\left[k_1 x - \frac{\sqrt{\varepsilon_1}}{\sqrt{\varepsilon_3}}\omega_1(t-t_2) - \omega_1 t_1\right]\left\{e^{\left[\frac{\sqrt{\varepsilon_1}}{\sqrt{|\varepsilon_2|}}\omega_1(t_2-t_1)\right]} - e^{-\left[\frac{\sqrt{\varepsilon_1}}{\sqrt{|\varepsilon_2|}}\omega_1(t_2-t_1)\right]}\right\}\right\}\\ &-\frac{\sqrt{\mu_1\mu_0}}{\sqrt{\varepsilon_3\varepsilon_0}}\left\{+\frac{1}{4}\left(\frac{\varepsilon_1}{\varepsilon_3}-\frac{\sqrt{\varepsilon_1}}{\sqrt{\varepsilon_3}}\right)\cos\left[k_1 x + \frac{\sqrt{\varepsilon_1}}{\sqrt{\varepsilon_3}}\omega_1(t-t_2) - \omega_1 t_1\right]\left\{e^{\left[\frac{\sqrt{\varepsilon_1}}{\sqrt{|\varepsilon_2|}}\omega_1(t_2-t_1)\right]} + e^{-\left[\frac{\sqrt{\varepsilon_1}}{\sqrt{|\varepsilon_2|}}\omega_1(t_2-t_1)\right]}\right\}\right.\\ &\left.+\frac{1}{4}\left(\frac{\varepsilon_1}{\sqrt{|\varepsilon_2|}\sqrt{\varepsilon_3}}+\frac{\sqrt{\varepsilon_1}\sqrt{|\varepsilon_2|}}{\varepsilon_3}\right)\sin\left[k_1 x + \frac{\sqrt{\varepsilon_1}}{\sqrt{\varepsilon_3}}\omega_1(t-t_2) - \omega_1 t_1\right]\left\{e^{\left[\frac{\sqrt{\varepsilon_1}}{\sqrt{|\varepsilon_2|}}\omega_1(t_2-t_1)\right]} - e^{-\left[\frac{\sqrt{\varepsilon_1}}{\sqrt{|\varepsilon_2|}}\omega_1(t_2-t_1)\right]}\right\}\right\}\end{aligned} \quad (S8.d)$$

## 4. E and H phase difference when $\varepsilon_2 < 0$

In this section we provide the theoretical calculations of the phase of the $E$ and $H$ field for times after the permittivity of the medium is changed from a positive ($\varepsilon_1 \geq 1$) to a negative value ($\varepsilon_2 < 0$) at $t = t_1$. As mentioned in the main text, for times $t > t_1$, the signal within the medium is no longer monochromatic as the FW and BW waves are frozen and grow/decay exponentially, respectively. However, here we use the notion of phase of the $E$ and $H$ fields for a better understanding of the spatial separation between their maxima and their zero values. Theoretical closed-form expressions of the phase of the $E$ field ($\varphi_E$) and $H$ field ($\varphi_E$) will be presented in this section.

As shown above, when the permittivity of the medium where a wave is traveling is changed from a positive ($\varepsilon_1 \geq 1$) to a negative value ($\varepsilon_2 < 0$) at $t = t_1$, the $E$ and $H$ fields within such medium can be defined using Eqs. S.7c and S.7e, respectively (Eqs. 1a and 1b from the main text, respectively). Then we can rewrite Eqs. S.7c and S.7e using the identity $A\sin(\theta) + B\cos(\theta) = C\sin(\theta + \varphi)$ with $C = \sqrt{A^2 + B^2}$, $\varphi = atan\left(\frac{B}{A}\right)$, as follows:

working with the E field defined by Eq. S.7c:

$$E_{2\_t1} = C\sin(\theta + \varphi_E) \quad \text{(S.9a)}$$

$$A_E = \frac{1}{2}\frac{\sqrt{\varepsilon_1}}{\sqrt{|\varepsilon_2|}} e^{-\left[\frac{\sqrt{\varepsilon_1}}{\sqrt{|\varepsilon_2|}}\omega_1(t-t_1)\right]} - \frac{1}{2}\frac{\sqrt{\varepsilon_1}}{\sqrt{|\varepsilon_2|}} e^{\left[\frac{\sqrt{\varepsilon_1}}{\sqrt{|\varepsilon_2|}}\omega_1(t-t_1)\right]} \quad \text{(S.9b)}$$

$$B_E = -\frac{1}{2}\frac{\varepsilon_1}{|\varepsilon_2|} e^{-\left[\frac{\sqrt{\varepsilon_1}}{\sqrt{|\varepsilon_2|}}\omega_1(t-t_1)\right]} - \frac{1}{2}\frac{\varepsilon_1}{|\varepsilon_2|} e^{\left[\frac{\sqrt{\varepsilon_1}}{\sqrt{|\varepsilon_2|}}\omega_1(t-t_1)\right]} \quad \text{(S.9c)}$$

$$C_E = \sqrt{A_E^2 + B_E^2} \quad \text{(S.9d)}$$

$$\theta = k_1 x - \omega_1 t_1 \quad \text{(S.9e)}$$

Then:

$$\varphi_E = atan\left(\frac{B_H}{A_H}\right) = atan\left(\frac{-\frac{1}{2}\frac{\varepsilon_1}{|\varepsilon_2|}F - \frac{1}{2}\frac{\varepsilon_1}{|\varepsilon_2|}G}{\frac{1}{2}\frac{\sqrt{\varepsilon_1}}{\sqrt{|\varepsilon_2|}}F - \frac{1}{2}\frac{\sqrt{\varepsilon_1}}{\sqrt{|\varepsilon_2|}}G}\right) \quad \text{(S.9f)}$$

$$F = e^{-\left[\frac{\sqrt{\varepsilon_1}}{\sqrt{|\varepsilon_2|}}\omega_1(t-t_1)\right]} \quad \text{(S.9g)}$$

$$G = e^{\left[\frac{\sqrt{\varepsilon_1}}{\sqrt{|\varepsilon_2|}}\omega_1(t-t_1)\right]} \quad \text{(S.9h)}$$

Or simply

$$\varphi_E = atan\left[\frac{\sqrt{\varepsilon_1}}{\sqrt{|\varepsilon_2|}}\left(\frac{-F-G}{F-G}\right)\right] \quad (S.9i)$$

We can then apply the same process for the $H$ field to calculate its phase $\varphi_H$ for times $t > t_1^+$. Rewriting Eq. S.7e we get:

$$H_{2\_t1} = C_H \sin(\theta + \varphi_H) \quad (S.10a)$$

$$A_H = \frac{1}{2}\frac{\sqrt{\varepsilon_0}}{\sqrt{\mu_1\mu_0}}\frac{\varepsilon_1}{\sqrt{|\varepsilon_2|}}e^{-\left[\frac{\sqrt{\varepsilon_1}}{\sqrt{|\varepsilon_2|}}\omega_1(t-t_1)\right]} - \frac{1}{2}\frac{\sqrt{\varepsilon_0}}{\sqrt{\mu_1\mu_0}}\frac{\varepsilon_1}{\sqrt{|\varepsilon_2|}}e^{\left[\frac{\sqrt{\varepsilon_1}}{\sqrt{|\varepsilon_2|}}\omega_1(t-t_1)\right]} \quad (S.10b)$$

$$B_H = \frac{1}{2}\frac{\sqrt{\varepsilon_1\varepsilon_0}}{\sqrt{\mu_1\mu_0}}e^{-\left[\frac{\sqrt{\varepsilon_1}}{\sqrt{|\varepsilon_2|}}\omega_1(t-t_1)\right]} + \frac{1}{2}\frac{\sqrt{\varepsilon_1\varepsilon_0}}{\sqrt{\mu_1\mu_0}}e^{\left[\frac{\sqrt{\varepsilon_1}}{\sqrt{|\varepsilon_2|}}\omega_1(t-t_1)\right]} \quad (S.10c)$$

$$C_H = \sqrt{A_H^2 + B_H^2} \quad (S.10d)$$

$$\theta = k_1 x - \omega_1 t_1 \quad (S.10e)$$

Then:

$$\varphi_H = atan\left(\frac{B_H}{A_H}\right) = atan\left(\frac{\frac{1}{2}\frac{\sqrt{\varepsilon_1\varepsilon_0}}{\sqrt{\mu_1\mu_0}}F + \frac{1}{2}\frac{\sqrt{\varepsilon_1\varepsilon_0}}{\sqrt{\mu_1\mu_0}}G}{\frac{1}{2}\frac{\sqrt{\varepsilon_0}}{\sqrt{\mu_1\mu_0}}\frac{\varepsilon_1}{\sqrt{|\varepsilon_2|}}F - \frac{1}{2}\frac{\sqrt{\varepsilon_0}}{\sqrt{\mu_1\mu_0}}\frac{\varepsilon_1}{\sqrt{|\varepsilon_2|}}G}\right) \quad (S.10f)$$

$$F = e^{-\left[\frac{\sqrt{\varepsilon_1}}{\sqrt{|\varepsilon_2|}}\omega_1(t-t_1)\right]} \quad (S.10g)$$

$$G = e^{\left[\frac{\sqrt{\varepsilon_1}}{\sqrt{|\varepsilon_2|}}\omega_1(t-t_1)\right]} \quad (S.10h)$$

Or simply

$$\varphi_H = atan\left[\frac{\sqrt{|\varepsilon_2|}}{\sqrt{\varepsilon_1}}\left(\frac{F+G}{F-G}\right)\right] \quad (S.10i)$$

From the expressions of $\varphi_E$ and $\varphi_H$ (Eqs. S.9i and S.10i, respectively), one can notice that their phase evolves as a function of time, as expected. Finally, we can write the phase difference as

$$\varphi_E - \varphi_H = atan\left[\frac{\sqrt{\varepsilon_1}}{\sqrt{|\varepsilon_2|}}\left(\frac{-F-G}{F-G}\right)\right] - atan\left[\frac{\sqrt{|\varepsilon_2|}}{\sqrt{\varepsilon_1}}\left(\frac{F+G}{F-G}\right)\right] \quad (S.11)$$

The expressions from Eqs. S.9i, S.10i and S.11 were then used to represent the results shown in Fig. 2c in the main text, noting that the phase difference $\varphi_E - \varphi_H$ will only be 90 degrees for times $t = \infty$. This is because when $t = \infty$, F and G in Eq. S.11 become 0 and $\infty$, respectively. Meaning that, in the limit when $t = \infty$ (or in the limit when $G = \infty$) $F = 0$ and the arguments of the atan () functions in S.11a become:

$$\lim_{G=\infty} \frac{\sqrt{\varepsilon_1}}{\sqrt{|\varepsilon_2|}} \left(\frac{-F-G}{F-G}\right) = \frac{\sqrt{\varepsilon_1}}{\sqrt{|\varepsilon_2|}} \tag{S.12a}$$

$$\lim_{G=\infty} \frac{\sqrt{|\varepsilon_2|}}{\sqrt{\varepsilon_1}} \left(\frac{F+G}{F-G}\right) = -\frac{\sqrt{|\varepsilon_2|}}{\sqrt{\varepsilon_1}} \tag{S.12b}$$

hence,

$$\varphi_E - \varphi_H = atan\left[\frac{\sqrt{\varepsilon_1}}{\sqrt{|\varepsilon_2|}}\right] - atan\left[-\frac{\sqrt{|\varepsilon_2|}}{\sqrt{\varepsilon_1}}\right] = \pi/2 \tag{S.12c}$$

Demonstrating a phase difference of 90 degrees, which occurs only when $t = \infty$, as mentioned in the main text.

## 5. Forward and backward waves results: single step function of permittivity.

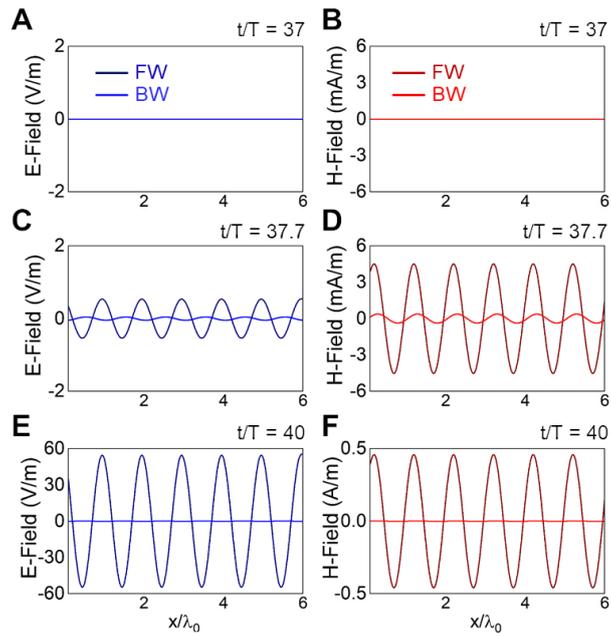

**Figure S2| Analytical FW and BW waves step function $\varepsilon_2 < 0$.** Electric (first column) and magnetic (second column) fields at different times when the permittivity is changed from $\varepsilon_1 = 1$ to $\varepsilon_2 = -20$ at $t_1 = 37.1T$. The FW and BW waves are plotted as dark and light lines for the electric (blue) and magnetic (red) fields, respectively

## 6. *Forward and backward waves results: square function of permittivity.*

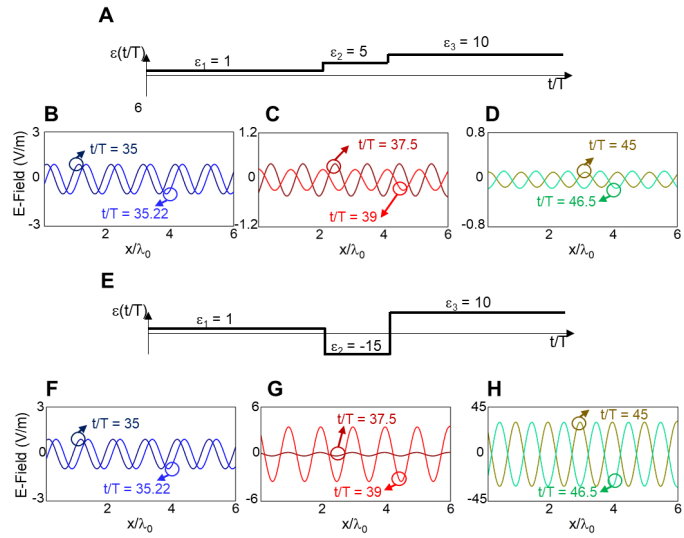

**Figure S3| Electric field distribution square function of ε. a-d** Positive-to-positive-to positive temporal slab. **b-d** Electric field distributions as a function of space (x) at t = 35T (i.e., before $t_1$, dark blue) and t = 35.22T (i.e., before $t_1$, light blue) **b**, at t = 37.5T (i.e., $t_1 < t < t_2$, dark red) and t = 39T (i.e., $t_1 < t < t_2$, light red) **c**, and at t = 45T (i.e., $t > t_2$, dark yellow) and t = 46.5T (i.e., $t > t_2$, light green) **d**. **e-h** Positive-to-negative-to positive temporal slab. **f-h** same as **b-d** but for the permittivity from **e**.

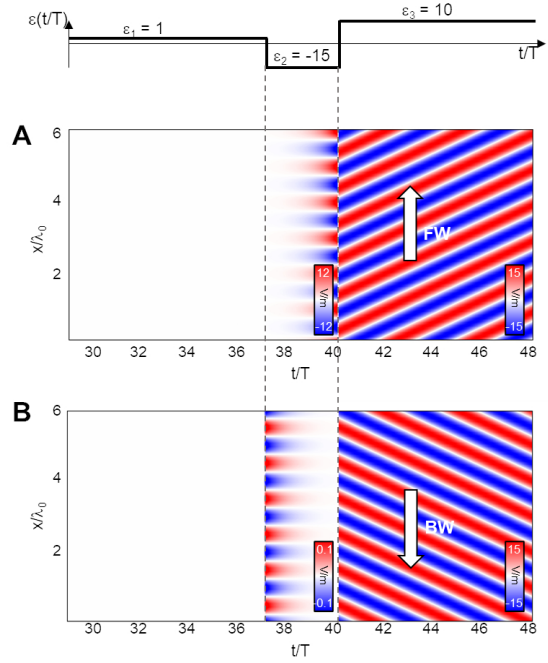

**Figure S4| Analytical FW and BW waves square function $\varepsilon_2 < 0$. a,b** Electric field distributions for the FW and BW waves, respectively, as a function of space and time when $\varepsilon$ is changed from $\varepsilon_1 = 1$ to a negative value $\varepsilon_2 = -15$ at $t_1 = 37T$ and then changed to $\varepsilon_3 = 10$ at $t_1 = 40T$. The time dependent $\varepsilon$ is shown on top of panel a.

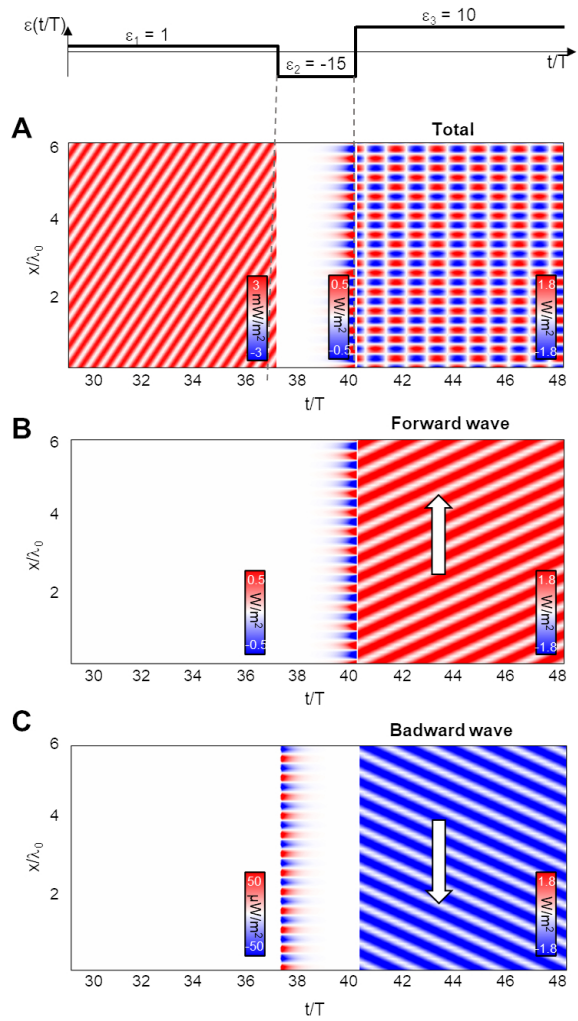

**Figure S5| Instantaneous Poynting vector square function $\varepsilon_2 < 0$. a,b,c** Instantaneous Poynting vector for the total, FW and BW waves, respectively, as a function of space and time when $\varepsilon$ is changed from $\varepsilon_1 = 1$ to a negative value $\varepsilon_2 = -15$ at $t_1 = 37T$ and then changed to $\varepsilon_3 = 10$ at $t_2 = 40T$. The time dependent $\varepsilon$ is shown on top of panel a.

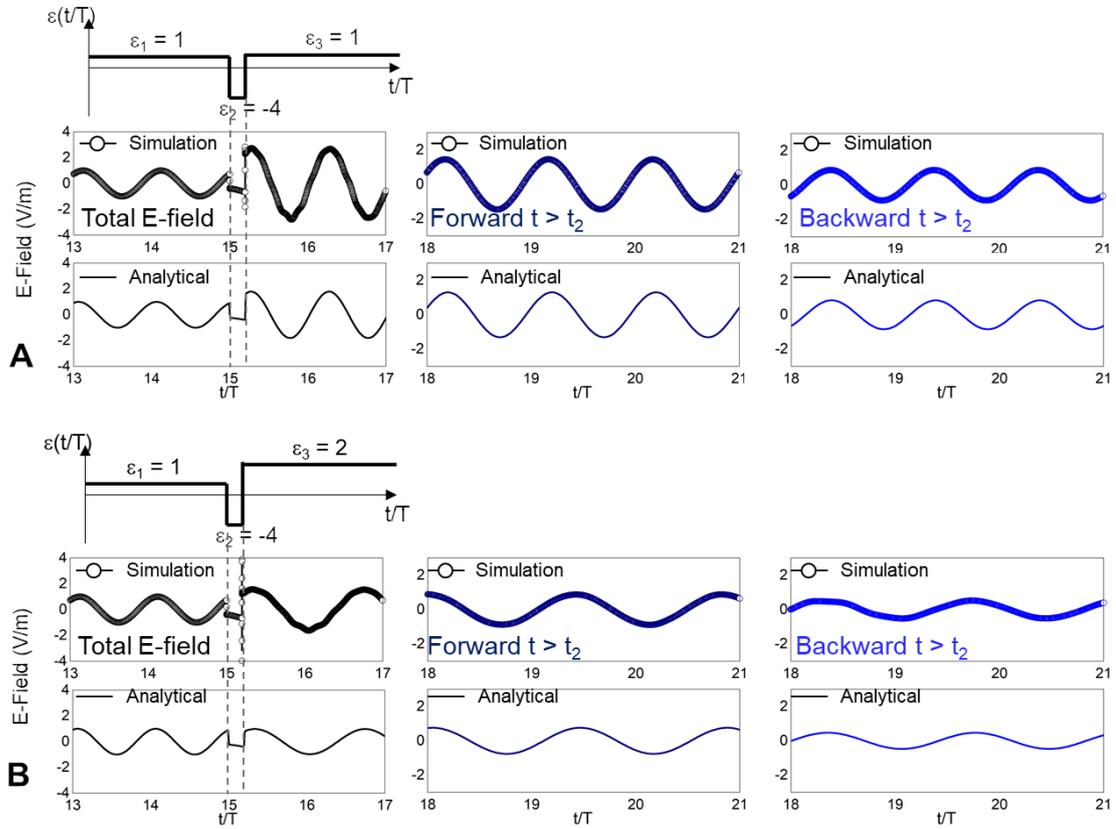

**Figure S6| Analytical and numerical results square function $\varepsilon_2 < 0$.** Simulation (circles) and numerical (solid lines) results of the total electric field (left column), forward (middle column) and backward (right column) distribution when the $\varepsilon$ is changed from $\varepsilon_1 = 1$ to: **a,** $\varepsilon_2 = -4$ and returned to $\varepsilon_3 = 1$, **b,** $\varepsilon_2 = -4$ and then to $\varepsilon_3 = 2$. In all the cases the first and second temporal boundaries are introduced at $t_1 = 15T$ and $t_2 = 15.2T$, respectively. The time dependent $\varepsilon$ is shown on top of each panel.

## 7. *Instantaneous Poynting vector: effect of $t_2 - t_1$.*

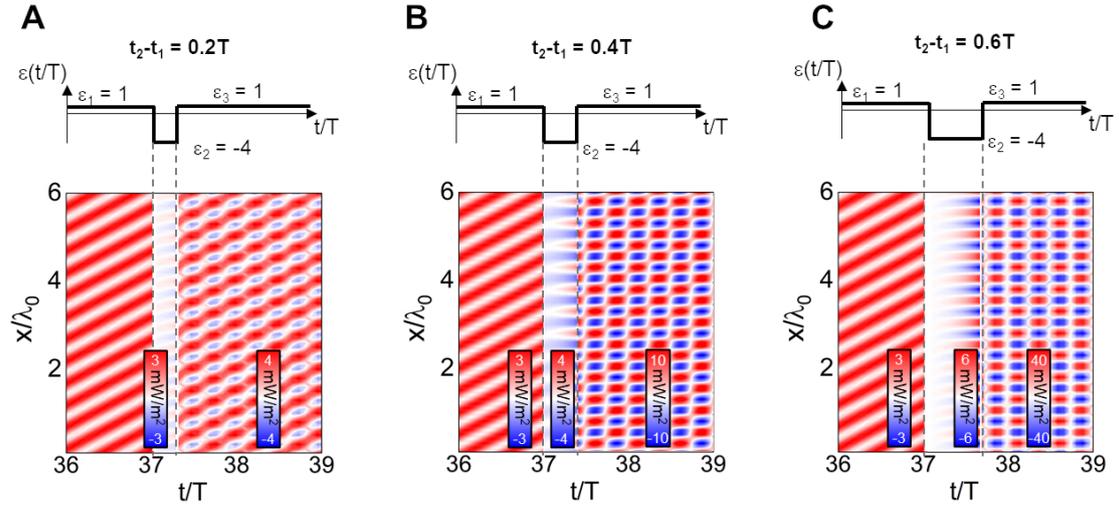

**Figure S7| Instantaneous Poynting vector square function $\varepsilon_2 < 0$.** Instantaneous Poynting vector for the total field as a function of space and time for the cases discussed in Fig. 4 of the main manuscript: **a,** $\varepsilon$ is changed from $\varepsilon_1 = 1$ to a negative value $\varepsilon_2 = -4$ at $t_1 = 37T$ and then changed to $\varepsilon_3 = 1$ at $t_2 = 37.2T$. **b,** $\varepsilon$ is changed from $\varepsilon_1 = 1$ to a negative value $\varepsilon_2 = -4$ at $t_1 = 37T$ and then changed to $\varepsilon_3 = 1$ at $t_2 = 37.4T$. **c,** $\varepsilon$ is changed from $\varepsilon_1 = 1$ to a negative value $\varepsilon_2 = -4$ at $t_1 = 37T$ and then changed to $\varepsilon_3 = 1$ at $t_2 = 37.6T$. The time dependent $\varepsilon$ is shown on top of each panels.

## 8. *Laplace transform derivation of electric field after temporal boundary at t = 0.*

Starting with the Maxwell equations

$$\frac{\partial E_y}{\partial x} = -\frac{\partial B_z}{\partial t} \tag{S.13a}$$

$$\frac{\partial H_z}{\partial x} = -\frac{\partial D_y}{\partial t} \tag{S.13b}$$

we apply the Laplace transform on the field quantities $E_y(x,t), D_y(x,t), H_z(x,t)$ and $B_z(x,t)$, resulting in

$$\frac{\partial \tilde{E}_y(x,s)}{\partial x} = -[s\tilde{B}_z(x,s) - B_z(x, t=0_-)] \tag{S.14a}$$

$$\frac{\partial \tilde{H}_z(x,s)}{\partial x} = -[s\tilde{D}_y(x,s) - D_y(x, t=0_-)] \tag{S.14b}$$

where $\tilde{E}_y(x,s)$, $\tilde{D}_y(x,s)$, $\tilde{H}_z(x,s)$ and $\tilde{B}_z(x,s)$ are the Laplace transforms of the quantities $E_y(x,t), D_y(x,t), H_z(x,t)$ and $B_z(x,t)$, respectively for $t \geq 0$. The terms $B_z(x,t=0_-)$ and $D_y(x,t=0_-)$ represent the initial values of these quantities at $t=0_-$. At the temporal boundary $t=0$, we have $B_z(x,t=0_+) = B_z(x,t=0_-)$ and $D_y(x,t=0_+) = D_y(x,t=0_-)$, leading to $H_{2z}(x,t=0_+) = H_{1z}(x,t=0_-)$ and $\varepsilon_2 E_{2y}(x,t=0_+) = \varepsilon_1 E_{1y}(x,t=0_-)$ where $\varepsilon_1 > 0$ and $\varepsilon_2 < 0$ are the relative permittivity of the medium before and after $t=0$, respectively, and $E_{1y}(x,t=0_-)$ and $E_{2y}(x,t=0_+)$ are the field values at $t=0_-$ and $t=0_+$, respectively. From Eq. (S.14), we obtain

$$\frac{\partial^2 \tilde{E}_{2y}(x,s)}{\partial x^2} = -\mu_0\varepsilon_0\varepsilon_2 s^2 \tilde{E}_{2y}(x,s) + \mu_0\varepsilon_0\varepsilon_2 s E_{2y}(x,t=0_+) = \mu_0 \frac{\partial H_z(x,t=0_+)}{\partial x} \tag{S.15}$$

Assuming the original electric and magnetic fields are given as $E_{1y}(x,t) = \cos(k_1 x - \omega_1 t)$ and $H_{1y}(x,t) = \sqrt{\frac{\varepsilon_1 \varepsilon_0}{\mu_0}} \cos(k_1 x - \omega_1 t)$, the solution to Eq. (S.15) can be written as

$$\tilde{E}_{2y}(x,s) = \frac{\varepsilon_1}{\varepsilon_2} \frac{s\cos(k_1 x) + \omega_1 \sin(k_1 x)}{s^2 + \omega_1^2 \frac{\varepsilon_1}{\varepsilon_2}} \tag{S.16}$$

Considering the fact that $\varepsilon_2 < 0$, which can be rewritten as $\varepsilon_2 = -|\varepsilon_2|$, we apply the inverse Laplace transform on Eq. (S.16) and we obtain

$$E_{2y}(x,t) = -\frac{\varepsilon_1}{|\varepsilon_2|}\cos(k_1 x)\cosh\left(\omega_1 t\sqrt{\frac{\varepsilon_1}{|\varepsilon_2|}}\right) - \sqrt{\frac{\varepsilon_1}{|\varepsilon_2|}}\sin(k_1 x)\sinh\left(\omega_1 t\sqrt{\frac{\varepsilon_1}{|\varepsilon_2|}}\right) \qquad (S.17)$$

which after simple mathematical manipulation, is written as

$$E_{2y}(x,t) = \begin{aligned}&\frac{1}{2}\left[-\frac{\varepsilon_1}{|\varepsilon_2|}\cos(k_1 x) - \sqrt{\frac{\varepsilon_1}{|\varepsilon_2|}}\sin(k_1 x)\right]e^{+\omega_1 t\sqrt{\frac{\varepsilon_1}{|\varepsilon_2|}}}\\&+\frac{1}{2}\left[-\frac{\varepsilon_1}{|\varepsilon_2|}\cos(k_1 x) + \sqrt{\frac{\varepsilon_1}{|\varepsilon_2|}}\sin(k_1 x)\right]e^{-\omega_1 t\sqrt{\frac{\varepsilon_1}{|\varepsilon_2|}}}\end{aligned} \qquad (S.18)$$

This expression is the same as Eq. (1a) in the main text and Eq. S.7c from this supplementary materials document when $t_1 = 0$.

## 9. Non-Foster structures with loss.

Before bringing loss into the picture, let us first take a closer look at the nondispersive time-varying scenario in the main manuscript, with an instantanenous polarization response parameterized by $P(t) = \varepsilon_0 \varepsilon(t) E(t)$ (no magnetic response is considered throughout), and consider a transition to some non-Foster medium with by $\varepsilon(t) < 0$, starting from vacuum for simplicity: it is clear that, in the picture of a smoothed ($C^1$-continuous, though arbitrarily abrupt) step-function change in $\varepsilon(t)$, we have a singularity at the epsilon-near-zero (ENZ) condition $\varepsilon(t) = 0$. With the plane wave propagation under study, and considering momentum ($k$) conservation, the wave equation can be written as $\frac{\partial^2(\varepsilon(t)E(t))}{\partial t^2} + (kc)^2 E(t) = 0$, which can be easily solved with a simple finite-difference scheme, in $t$ only. This allows us to observe this singular transition numerically, as shown in the plot of $E(t, x = 0)$ below, where our time-dependent epsilon changes from $\varepsilon_1 = 1$ to $\varepsilon_2 = -4$ with a transition time of $T/10^4$. As the time step $\Delta_t$ decreases, we get closer to the ENZ condition in our discrete time domain and, consequently, from the continuity of $D$, $E$ becomes larger and larger (see the zoom-in view of Fig. S8b, where the dashed black line represents $\varepsilon(t)$, modeled with the tanh() function). (Actually, if $\Delta_t$ were so coarse as to not sample any point across the transition ramp of $\varepsilon(t)$, the resulting $E$ would also behave as a step-function at $t = 0$, with no spikes, just like in the analytical solution.)

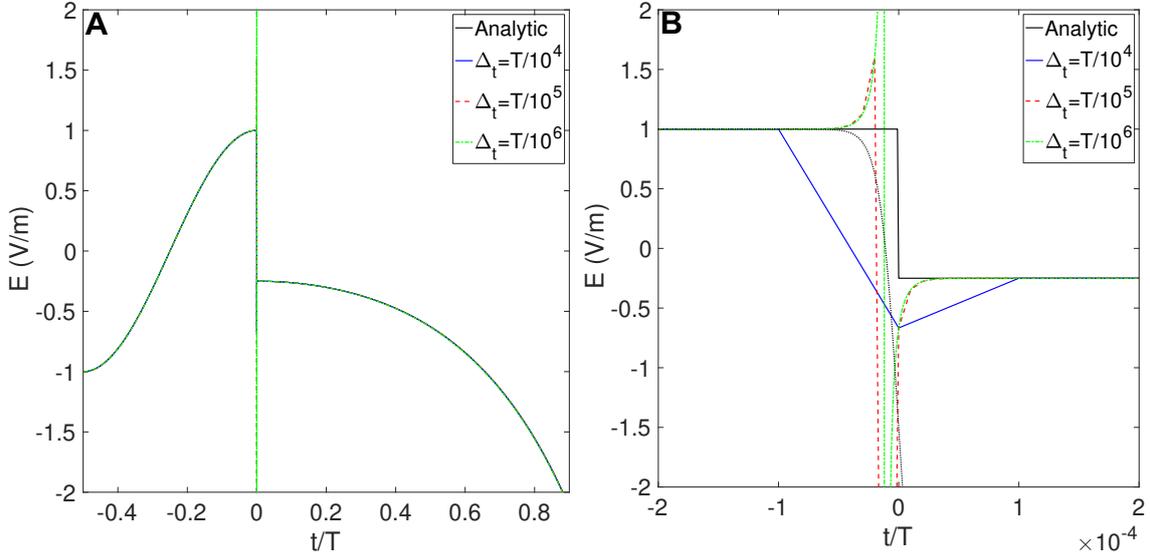

**Figure S8| Temporal evolution of electric field in lossless scenario. a** Comparison of numerical and analytic electric field vs. time at $z = 0$ with several sizes of the time step, for a transition from $\varepsilon_1 = 1$ to $\varepsilon_2 = -4$ (transition time of $T/10^4$). **b** Zoom-in view of **a** showing the divergent behavior of $E(t)$ when $\varepsilon(t) = 0$.

The analytic solution for an ideal (discontinuous) step-function is also included (black solid line) for comparison, and directly shows us how to circumvent the problem: using a circuit analogy (or more rigorously, a lumped-element transmission line) where our dielectric susceptibility $\chi$ is described by a per-unit-length series capacitance $C(t) = \varepsilon_0 \chi(t)$, we can think of a non-singular transition as the closing of an electric switch that connects this capacitor $C(t)$ in parallel with $\varepsilon_0$. Although the dielectric polarization (charge) is driven by the electric field (voltage), in the following we will be using several time-varying (switched) toy-model circuits (i.e., no wave propagation involved) driven by a continuous-wave (CW) current source, in order to get some insight into the actual problem of wave propagation. (We will therefore use without distinction $C(t)$ in (F) to denote capacitance in a circuit, and $C(t)$ in (F/m) to denote capacitance per distance in a transmission line.) This stems from the understanding that, in the latter, current, and not voltage, preserves continuity across the jump in $\varepsilon(t)$: in any nonmagnetic time-varying material, the continuity of $B$ implies the continuity of $H$ and thus of its curl, so not only do we have a continuous charge $D$ but also a continuous current $\frac{dD}{dt}$.

In this picture, in the vicinity of $t = 0$ the nondispersive scenario studied in Fig. S8a has many similarities with the current-driven circuit of Fig. S9a, where we have further allowed for a resistor in series with our smoothly-varying capacitor which models any parasitic losses. The presence of this resistor imposes an upper bound on the voltage across our time-varying capacitor $C(t)$. If we first assume an adiabatic variation of $C(t)$, we can drop the term $\frac{dC(t)}{dt} v_1(t)$ in $i_1(t)$ and write

$$v_1(t) = \frac{1}{1+\frac{C(t)}{\varepsilon_0}+i\omega RC(t)} \frac{1}{i\omega \varepsilon_0} i_s(t) \tag{S.19}$$

with $\omega$ the oscillation frequency of the source $i_s(t)$. In the limit of $C(t) = -\varepsilon_0$ (or equivalently, $\varepsilon(t) = 0$), we have $v_1(t) = -\frac{i_s(t)}{R(\omega \varepsilon_0)^2}$, which indeed only diverges as $R \to 0$. With this in mind, let us skip the numerical solution (otherwise trivial) of the currents and voltages in this circuit across a continuous step-function variation of $C(t)$, and go back to our original propagation problem in Fig. S8. But now we will include the role of the resistor in our new polarization response $R\frac{dP(t)}{dt} + \frac{P(t)}{C(t)} = E(t)$, which under a slowly-varying $C(t)$ and assuming CW can be rewritten as $P(t) = \frac{C(t)}{1+i\omega RC(t)} E(t)$, i.e., $\varepsilon(t) = 1 + \frac{C(t)/\epsilon_0}{1+i\omega RC(t)}$. A rigorous mathematical treatment of this response—valid for any variation of $C(t)$—is considerably more convoluted (see [4] for a detailed explanation), but we can nonetheless still write a wave equation for $P$ as

$$\left(\frac{\partial^2}{\partial t^2} + (kc)^2\right)\left(R\frac{\partial P(t)}{\partial t} + \frac{P(t)}{C(t)}\right) = -\frac{1}{\varepsilon_0}\frac{\partial^2 P(t)}{\partial t^2} \qquad (S.20)$$

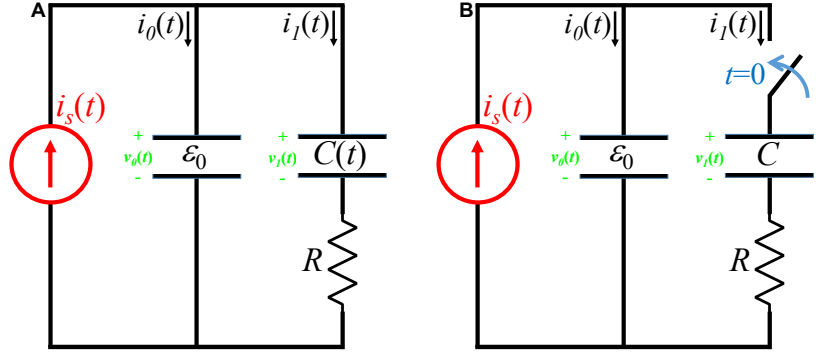

**Figure S9| CW current-driven circuit models capturing some of the physics of our plane-wave time interface.** **a** Continuous and **b** abrupt step-function transition of $\varepsilon(t)$.

In Fig. S10 we start from the same temporal variation of the dielectric function of Fig. S8, but include the effect of several values of the resistor $R = (-1, -0.1, -0.01)$, which parameterize an effective jump in the (now dispersive) dielectric function from $\varepsilon(\omega) \approx 1$ to $\varepsilon(\omega) \approx (-3.9655 - 0.4141i, -3.9997 - 0.0417i, -4.0000 - 0.0042i)$, respectively, where $\omega$ is the initial frequency. Importantly, we stress out that $R \leq 0$ initially. In order to understand this, let us move back to the circuit model: $C(t) \leq 0$ as soon as $\varepsilon(t) \leq 1$, and we thus need a negative resistor in order to keep stability. However, the condition for stability reaches a branch cut at $|C(t)| = \varepsilon_0$, meaning we need to flip the sign of the resistor (from negative to positive) right when $C(t)$ crosses $-\varepsilon_0$ in order to avoid instability (an interesting discussion on the topic can be found in [5]). In our propagation problem of Fig. S10, by flipping the sign of the resistor we are in actuality avoiding one of two possible instabilities, but the other one is precisely the amplification mechanism we are seeking (further details will be presented below). These simulation results show how the bound on $E$ when $C(t) = -\varepsilon_0$ decreases with $|R|$, so the value of $\Delta_t$ is now irrelevant, as long as it is small enough to track the velocity of the polarization response across the transition (we choose $\Delta_t = T/10^6$ in this case).

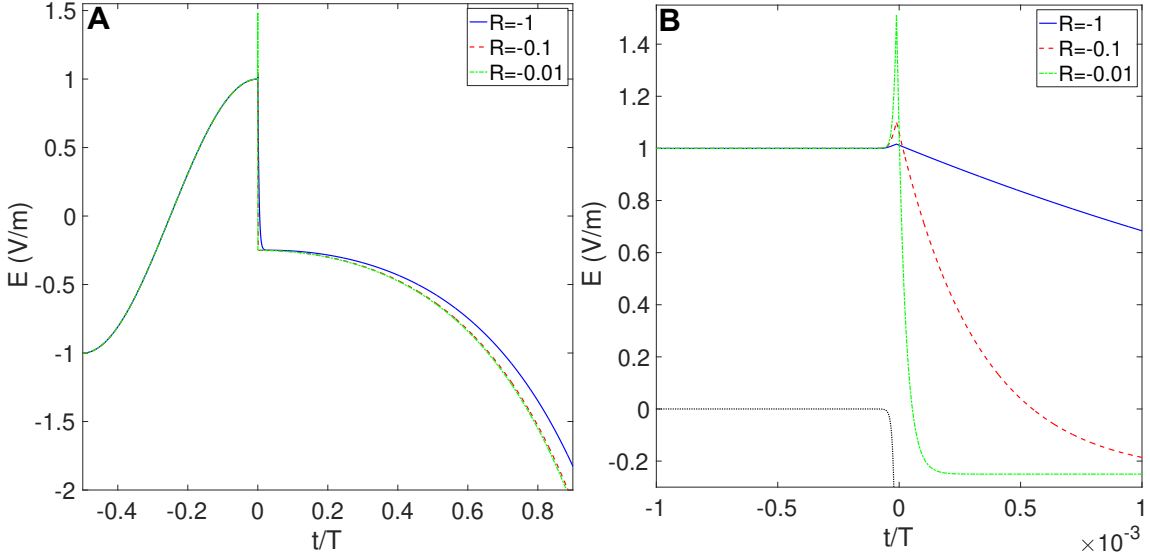

**Figure S10| Temporal evolution of electric field in lossy scenario of Fig. S9a. a** Numerical solution of the electric field vs. time at $z = 0$ when $C(t)$ changes from 0 to $-5\varepsilon_0$, considering several values of $R$ (equivalent to $\varepsilon_2 = -4$ in the limit of $R = 0$). **b** Zoom-in view of **a** showing the bounded behavior of $E(t)$ when $C(t) = -\varepsilon_0$, and how a larger $R$ also leads to slower responses (the dashed black line represents the transition of $\frac{C(t)}{\varepsilon_0}$ from 0 to $-5$).

### 9.1 Study of Stability.

Having clarified the physics at play when $C(t)$ varies continuously, we will hereinafter use electric switches modeling discontinuous step-function transitions, as shown in Fig. S9b. We analyse the transient response of this toy-model circuit for $t > 0$ with the help of the Laplace transform $\mathcal{LT}$ and write, for a harmonic source $\tilde{I}_s(s) = \frac{1}{s - i\omega}$:

$$\tilde{V}_0(s) = \frac{1+sRC}{s(\varepsilon_0 + C + sRC\varepsilon_0)} \left( \tilde{I}_s(s) + \varepsilon_0 v_0^- \right) \tag{S.21a}$$

$$\mathcal{LT}^{-1}: \quad v_0(t \geq 0) = -\frac{1+i\tau\omega}{\kappa\omega} e^{i\omega t} + \frac{C}{C+\varepsilon_0} \frac{i\tau + \kappa v_0^-}{\kappa} e^{-\frac{t}{\frac{C}{C+\varepsilon_0}\tau}} + v_{DC} \tag{S.21b}$$

with $\kappa = -i(C + \varepsilon_0) + \varepsilon_0 \tau \omega$, $\tau = RC$ the usual time constant, and $v_0^- \equiv v_0(t = 0^-)$. We note that, in our circuit to transmission-line equivalence, $v_0(t)$ represents the electric field $E$. Now, besides a DC component that is irrelevant for our discussion, we have a driven oscillating term $e^{i\omega t}$ plus a transient response with time constant $\frac{C}{C+\varepsilon_0}$. For a non-Foster capacitor and assuming $C < -\varepsilon_0$, we need $R \geq 0$ in order to place this pole in the left-half of the $s$-plane and bring the system to a stable point of operation.

The situation is more subtle when we consider propagation. From momentum conservation, we obtain the new frequencies for $t > 0$, $\omega_n$, by solving $\omega = \omega_n \sqrt{1 + \frac{C/\varepsilon_0}{1+i\omega_n RC}}$. There are three solutions to this transcendental equation:

$$\omega_n = \frac{1}{3\tau}\left(i(C+1) + 2^{1/3}e^{i(2n+1)\frac{\pi}{3}}\frac{K_1}{K_2} - 2^{-1/3}e^{-i(2n+1)\frac{\pi}{3}}K_2\right), \quad n=0,1,2, \tag{S.22}$$

where we define

$$K_0 = -2i\left(\frac{C}{\varepsilon_0}+1\right)^3 + 9i\left(\frac{C}{\varepsilon_0}-2\right)(\tau\omega)^2, \quad K_1 = \left(\frac{C}{\varepsilon_0}+1\right)^2 - 3(\tau\omega)^2, \quad K_2 = \left(K_0 + \sqrt{4K_1^3 + K_0^2}\right)^{1/3}$$

(S.23)

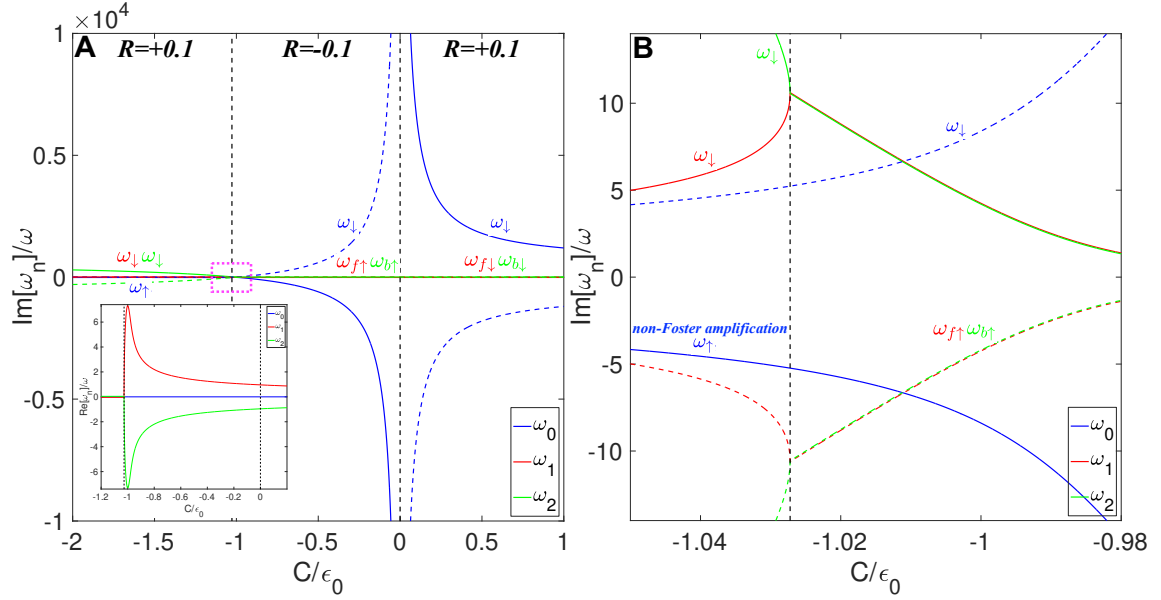

**Figure S11| Eigenfrequencies from Eq. S.22.** Normalized to the initial $\omega$, vs. $\frac{C}{\varepsilon_0}$, with the resistor fixed at $\pm 0.1(\Omega \cdot m)$. **a** Imaginary part (the inset shows the real part). **b** Zoom-in plot of **a** near $\frac{C}{\varepsilon_0} \to -1$ (dashed pink rectangle). Although not distinguishable with this scale, in the rightmost region the imaginary part of $\omega_{f\downarrow}$ and $\omega_{b\downarrow}$ increases with $C > 0$.

If we fix $R = \pm 0.1(\Omega \cdot m)$ and plot these solution frequencies as a function of $C$ (Fig. S11a, solid and dash lines lines for $R = +0.1$ and $R = -0.1$, respectively), one can observe that, as expected, if $C > 0$ and $R > 0$ (solid lines), the loci of the three solutions is in the upper-half of the complex $\omega$-plane, i.e., we have temporal decay: $\omega_\downarrow$ represents a purely-imaginary (overdamped) eigenfrequency and carries no propagation, whereas $\omega_{f\downarrow}$ and $\omega_{b\downarrow}$, underdamped, propagate forward and backward, respectively (the real part of these frequencies is shown in the inset of Fig. S11a). Once $C$ decreases to negative values, we need a negative resistor (dashed lines) in order for the purely-imaginary solution to stay stable; however, the propagating modes are now unstable, but the magnitude of this stability is controllable and can be harnessed precisely to achieve amplification. If one keeps increasing the negative magnitude of $C(t)$, another branch cut is found when the term $4K_1^3 + K_0^2$ in $K_2$ crosses zero, which in this example happens at $\frac{C}{\varepsilon_0} \approx -1.0272$ (unlike in the circuit model, $\frac{C}{\varepsilon_0} \to -1$ only in the limit of $R \to 0$). Therefore, once we cross this point (see Fig. S11b for a zoom-in view) we need $R > 0$ again (solid lines) to discard the problematic pole (dashed green lines in the leftmost region, separated by a dashed black line). We thus have a very rapidly decaying mode (green) plus a conjugate pair of decaying (red) and growing (blue) modes, the three of them purely imaginary (no propagation).

In Fig. S12a we fix $\frac{C}{\varepsilon_0} = -5$ and show the same parametric curves as a function of $R$, which reveal the discontinuity in the stability behaviour of our system at $R = 0$. As mentioned above, given that this value of $C$ satisfies $C < -\varepsilon_0$, the three frequencies are purely imaginary and the corresponding permittivities $\varepsilon(\omega_n)$ are purely real and negative (Fig. S12b), with the rapidly decaying mode (green line) very close to a perfect ENZ condition.

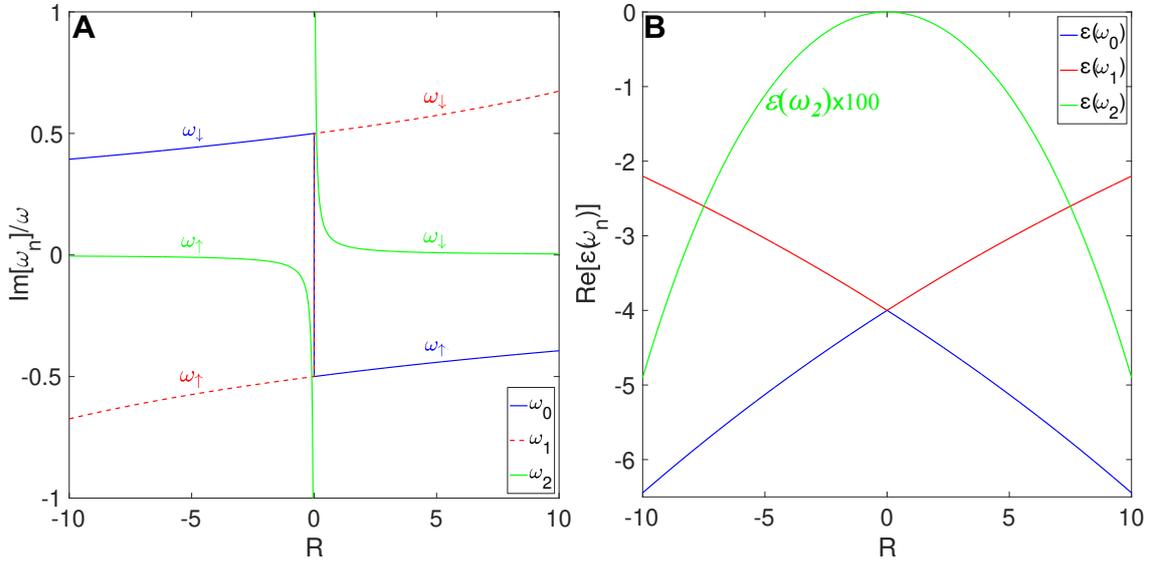

**Figure S12| Eigenfrequencies from Eq. S.22. a** Imaginary part (normalized to the initial $\omega$) vs. $R$, with $\frac{C}{\varepsilon_0} = -5$. **b** Corresponding values of the dielectric function at these frequencies ($\varepsilon(\omega_2) \approx 0$, so it is scaled by a factor of 100).

### 9.2 Amplitudes of the waves scattered at $t=t_1$.

Whereas in the temporal interface of a nondispersive dielectric function the needed boundary conditions (BCs) are the continuity of $D$ and $B$ ($H$ if $\mu = 1$)—$E$ and $P$ being discontinuous—, we now have three scattered waves, as shown before, so we need an extra BC. From Fig. S9b, it is straightforward to see that, when we close the switch at $t_1 = 0$, due to the presence of the resistor there is no instantaneous transfer of charge from $\varepsilon_0$ to $C$, so $v_1^+ = v_1^- = 0$. It is the voltage across the resistor ($v_R^+ = R i_1^+$) that equals $v_0^+ = v_0^-$: translating this behavior into the wave problem, $E$ and $P$ are now continuous, with the step discontinuity showing up in $\frac{dP^+}{dt} = \frac{E^-}{R}$. Following the recipe of [6], if we write

$$E(t \leq 0) = e^{-ikz}e^{i\omega t}, \qquad E(t \geq 0) = e^{-ikz}\sum_{n=0}^{2} a_n e^{i\omega_n t}, \qquad (S.24)$$

with $a_n$ the unknown complex coefficients in this superposition, we arrive at the matrix system of equations

$$\begin{bmatrix} 1 & 1 & 1 \\ \varepsilon(\omega_0) & \varepsilon(\omega_1) & \varepsilon(\omega_2) \\ \sqrt{\varepsilon(\omega_0)} & \sqrt{\varepsilon(\omega_1)} & \sqrt{\varepsilon(\omega_2)} \end{bmatrix} \begin{bmatrix} a_0 \\ a_1 \\ a_2 \end{bmatrix} = \begin{bmatrix} 1 \\ \varepsilon^- \\ \sqrt{\varepsilon^-} \end{bmatrix}, \qquad (S.25)$$

where the 1st, 2nd and 3rd rows represent the BCs for $E$, $D$ and $H$, respectively, at $t = 0$. Note also that the transition from air exemplified in Fig. S9b specializes this system to $\varepsilon^- = 1$. Fig. S13a depicts $|a_n|$ when $C$ is swept as in Fig. S11, with $R = \pm 0.1$ (the magnitude of these coefficients is independent of the sign of $R$), showing (i) how for $C = 0$ only $\omega_{f\downarrow}$ ($\omega_f$ in this limit case) survives ($a_n = 1$), commensurate with a parallel open-circuit; and (ii) how the magnitude of $a_1$ and $a_2$ diverge as we approach the leftmost branch cut and, more importantly, as we reduce $|R|$ and this point tends to $\frac{C}{\varepsilon_0} = -1$, $a_0$, bounded by $R$, also diverges. This should come at no surprise as, in the nondispersive scenario, $E^+ = \frac{D^-}{\varepsilon_{ENZ}}$ will tend to diverge. In Fig. S13b we repeat the temporal interface of Fig. S10b, but with a switch or ideal (not smooth) step function: the results are essentially the same as before, except for the disappearance of the spikes produced by $C(t) = 0$. As expected, in the limit $R \to 0$ we recover the nondispersive solution (black dashed line).

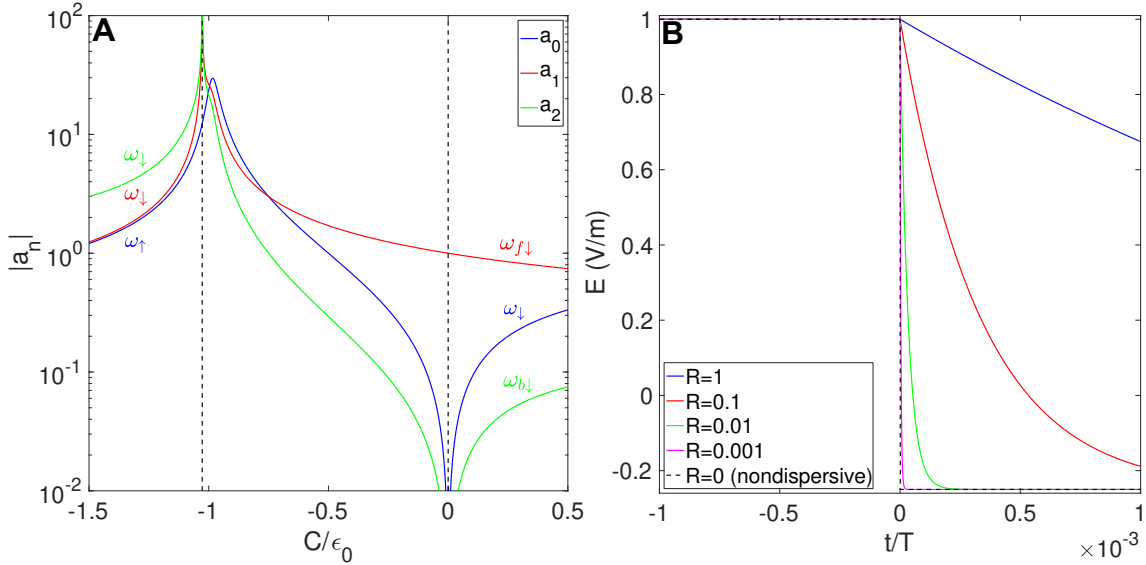

**Figure S13| Scattering coefficients and temporal evolution of electric field in lossy scenario of Fig. S2b. a** Magnitude of the coefficients of the temporally-scattered waves vs. $\frac{C}{\varepsilon_0}$, with the resistor fixed at $+0.1 (\Omega \cdot m)$. **b** Analytic and numerical (perfectly overlapped) solution of the electric field vs. time at $z = 0$ when the switch in Fig. S9b is closed at $t_1 = 0$, with $C = -5\varepsilon_0$ and several values of $R$ (analogous to Fig. S10b).

### 9.3 "Unfreezing" the forward and backward waves at $t=t_2$.

In the nondispersive scenario, once the metastructure is operating in the non-Foster amplification regime (following e.g. our transition from $\varepsilon^- = 1$ (air) to $\varepsilon^+ = -4$), we can revert the temporal change in the dielectric function and recover two amplified (forward- and backward-propagating) waves, as shown in the main manuscript. This process can be modeled by the time-varying circuit in Fig. S9a and becomes nondispersive in the limit $R \to 0$, but the use of switches might be more realistic in a real-life experiment (see e.g. the loaded parallel-plate waveguide proposed in Fig. 5 in the main text). In such case, it is clear that opening the switch in Fig. S9b does not reproduce the reverse transition. Even without the resistor, from the continuity of $D$ we need an (instantaneous) transfer of all the charge (polarization $P$) in the capacitor $C$ back to $\varepsilon_0$: opening the switch would leave the capacitor charged, though disconnected from the circuit and thus irresponsive to the driving electric field. One possible alternative would be the circuit illustrated in Fig. S14a: assuming the non-Foster capacitor $C$ has been charging up since $t_1$ ($t_1 = 0$ in our examples), we can think of connecting, at some instant $t_2 > t_1$, a new capacitor of the same magnitude and opposite sign that exactly cancels out the effect of $C < 0$. Resorting again to the Laplace transform, the transient voltages become

$$\tilde{V}_l(s) = \frac{\tilde{I}_s(s) + \varepsilon_0 v_0^- + (1+lsR\varepsilon_0)Cv_1^-}{s\varepsilon_0}, \quad l = 0,1 \tag{S.26a}$$

$$\mathcal{LT}^{-1}: \quad v_l(t' \geq 0) = i\frac{1-e^{i\omega t'}}{\varepsilon_0 \omega} + lRCv_1^- \delta(t') + \frac{\varepsilon_0 v_0^- + Cv_1^-}{\varepsilon_0}, \tag{S.26b}$$

with $t' = t - t_2$, and where the term $\frac{\varepsilon_0 v_0^- + Cv_1^-}{\varepsilon_0}$ delivers exactly what we are looking for: the effective transfer of charge from $C$ to $\varepsilon_0$. We also notice, however (and unless $R = 0$), the presence of a Dirac $\delta$ term in $v_1$, associated with mutually-cancelling instantaneous spikes of charge in $+C$ and $-C$ (their currents will thus contain terms in $\delta$ and $\frac{d\delta}{dt}$). This can be easily fixed in practice by adding a small resistor (negative, for similar reasons as above) to the newly-connected capacitor: in Fig. S14b we resume our case study example and switch our medium back to vacuum at $t_2 = T$, considering several values of this additional resistor. The numerical and analytic solutions—we omit these here for brevity, but they can be obtained in analogous fashion to Eqs. S.22,25—overlap perfectly and show how (i) $P(t > t_2)$ rapidly decays to zero (faster as we reduce the magnitude of the additional resistor) and (ii) how in the limit (i.e., no additional resistor) this transition tends to the described $\delta(t - t_2)$ terms in $P_{\pm C}$.

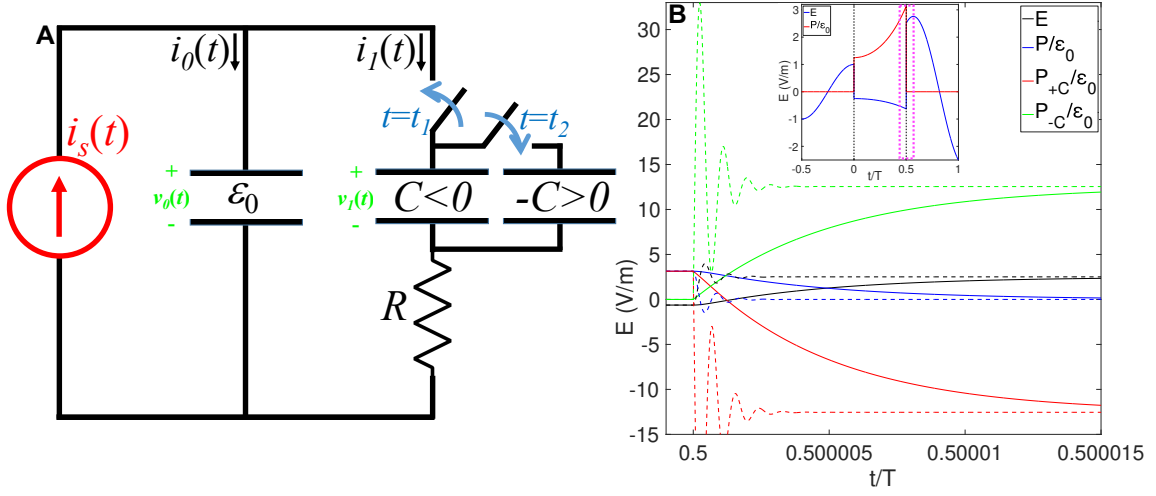

**Figure S14| Circuit toy model of the two transitions in the dielectric function at $t_1$ and $t_2$. a** Switched-circuit schematic. **b** Detailed view vs. time of the resulting electric field and the polarization in the capacitors around the transition at $t = t_2$, when using this circuit (transmission line) model to describe the polarization response of our time-varying medium, with $R = 10^{-4}$ $\Omega \cdot$ m. Solid (dashed) lines correspond to an additional negative resistor of $-10^{-4}$ ($-10^{-6}$) $\Omega \cdot$ m. The inset shows the entire simulation including the two temporal interfaces (this zoom-in region around the transition at $t = t_2$ falls inside the dashed pink rectangle).

In addition, we can also envision the alternative setup laid out in Fig. S15a, where an entire RC branch is connected at $t = t_2$. Let us skip for a moment the switches marked in pink color in the figure, and assume the new RC branch has $-C$ and $-R$. Similarly as in Fig. S14a, we have

$$\tilde{V}_0(s) = \frac{(1+sRC)\tilde{I}_s(s) + \varepsilon_0 v_0^- + (1+sR\varepsilon_0)Cv_1^-}{s\varepsilon_0(1+sRC)} \tag{S.27a}$$

$$\mathcal{LT}^{-1}: \quad v_0(t' \geq 0) = i\frac{1-e^{i\omega t'}}{\varepsilon_0 \omega} + \frac{C}{\varepsilon_0}v_1^- e^{-\frac{t'}{\tau}} + \frac{\varepsilon_0 v_0^- + Cv_1^-}{\varepsilon_0}. \tag{S.27b}$$

Just like in Eq. S.26, we achieve a quasi-instantaneous transfer of charge from $C$ to $\varepsilon_0$, but with no spikes of charge (nor current) involved. Moreover, now we have an additional term in the form of a purely-real pole ($e^{-t'/\tau}$), so we need $RC > 0$ in order to ensure stability. This means that we have to flip the sign of the resistor in the non-Foster branch, which becomes negative after $t = t_2$ (hence the switches marked in pink color in Fig. S15a), and as a result the resistor in the new $RC$ branch must be positive. In Fig. S15b we repeat the numerical experiment of Fig. S14b with this new setup and obtain very similar results, except for the different nature of the transient modes involved.

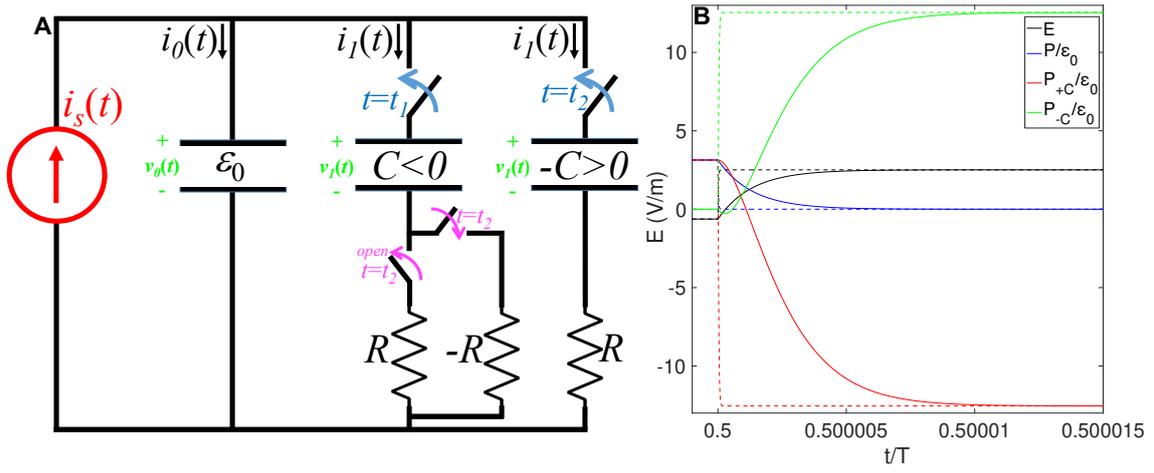

**Figure S15| Alternative circuit model of the two transitions of the dielectric function in Fig. S14. a** Switched-circuit schematic. **b** Same as in Fig. S14b, with $R = 10^{-4}$ $\Omega \cdot m$ (solid curves) and $R = 10^{-6}$ $\Omega \cdot m$ (dashed curves).

## 10. Impossibility of rapid temporal variation of permittivity (from positive to negative values) in Drude passive media.

Consider a homogeneous unbounded lossless passive medium with the Drude dispersion

$$\varepsilon = \varepsilon_0 \left(1 - \frac{\omega_p^2}{\omega^2}\right) \tag{S.28}$$

If we have a monochromatic wave with radian frequency $\omega_1$ such that $\omega_1 > \omega_{p1}$, propagating in such medium, the permittivity of the medium for this signal is

$$\varepsilon_1 = \varepsilon_0 \left(1 - \frac{\omega_{p1}^2}{\omega_1^2}\right) \tag{S.29}$$

which is a positive quantity. Now let us assume we want the permittivity of this medium to temporally change rapidly from a positive to a negative value by increasing its plasma frequency to $\omega_{p2}$ higher than $\omega_1$. However, when we change the value of the plasma frequency, since the wave momentum should stay conserved the signal frequency does change in order to satisfy the condition

$$\omega_1 \sqrt{\mu_1 \varepsilon_1} = \omega_2 \sqrt{\mu_1 \varepsilon_2} \tag{S.30}$$

This results in

$$\omega_1^2 \left(1 - \frac{\omega_{p1}^2}{\omega_1^2}\right) = \omega_2^2 \left(1 - \frac{\omega_{p2}^2}{\omega_2^2}\right) \tag{S.31}$$

Implying that

$$\omega_2^2 = \omega_{p2}^2 + \omega_1^2 - \omega_{p1}^2 > \omega_{p2}^2 \tag{S.32}$$

This results in $\varepsilon_2 > 0$. Therefore, it is impossible to make such temporal change of permittivity in such passive media. Consequently, we need to utilize the non-Foster structure to force and keep $\varepsilon_2 < 0$

## 11. Index of Supplementary Movies

Supplementary Movie 1: Positive-to-positive-to-positive change of permittivity.

Supplementary Movie 2: Positive-to-negative-to-positive change of permittivity.